\documentclass[aps, prd, superscriptaddress, nofootinbib, twocolumn]{revtex4-2}

\usepackage{amsmath}
\usepackage{graphicx}
\usepackage{dcolumn}
\usepackage{bm}
\usepackage{wasysym}
\usepackage{hyperref}
\usepackage[mathlines]{lineno}
\usepackage{xcolor}

\usepackage{gensymb}
\usepackage{enumitem}
\usepackage{multirow}
\usepackage{booktabs}
\usepackage{array}
\newcolumntype{P}[1]{>{\centering\arraybackslash}p{#1}}
\usepackage{amssymb}
\usepackage{adjustbox}
\usepackage{orcidlink}
\usepackage{pifont}
\usepackage{ulem}

\newcommand*{\defeq}{\mathrel{\vcenter{\baselineskip0.5ex \lineskiplimit0pt
                     \hbox{\scriptsize.}\hbox{\scriptsize.}}}
                     =}

\newcommand{\iu}{\mathrm{i}\mkern1mu}
\newcommand{\du}{\mathrm{d}}

\newcounter{count}
\definecolor{mygreen}{rgb}{0.1, 0.6, 0.2}

\definecolor{ABlue}{rgb}{0.0,0.7,0.8}

\begin{document}

\title{Lessons from binary dynamics of inspiralling equal-mass boson-star mergers} 

\author{Tamara Evstafyeva \orcidlink{0000-0002-2818-701X}}
\email{tevstafyeva@perimeterinstitute.ca}
\affiliation{Perimeter Institute for Theoretical Physics,
Waterloo, Ontario N2L 2Y5, Canada}

\author{Antonia Seifert \orcidlink{0009-0005-9892-3667}}
\email{aseifert@perimeterinstitute.ca}
\affiliation{Perimeter Institute for Theoretical Physics,
Waterloo, Ontario N2L 2Y5, Canada}
\affiliation{Department of Physics and Astronomy, University of Waterloo, Waterloo, Ontario, N2L 3G1, Canada}

\author{Ulrich Sperhake 
\orcidlink{0000-0002-3134-7088}}
\email{U.Sperhake@damtp.cam.ac.uk}
\affiliation{DAMTP, Centre for Mathematical Sciences, University of Cambridge, Wilberforce Road, Cambridge CB3 0WA, UK}
\affiliation{Department of Physics and Astronomy, Johns Hopkins University, 3400 North Charles Street, Baltimore, Maryland 21218, USA}
\affiliation{{TAPIR 350-17, Caltech, 1200 E. California Boulevard, Pasadena, California 91125, USA}}

\author{Christopher J.\ Moore}
\affiliation{Institute of Astronomy, University of Cambridge, Madingley Road, Cambridge, CB3 0HA, UK}
\affiliation{DAMTP, Centre for Mathematical Sciences, University of Cambridge, Wilberforce Road, Cambridge CB3 0WA, UK}
\affiliation{Kavli Institute for Cosmology, University of Cambridge, Madingley Road, Cambridge, CB3 0HA, UK}

\author{Tamanna Jain}
\affiliation{%
LPENS, Département de physique, Ecole normale supérieure, Université PSL, Sorbonne Université, Université Paris Cité, CNRS, 75005 Paris
}
\date{\today}

\begin{abstract}
    We explore the gravitational-wave phenomenology of equal-mass inspiralling boson-star binaries using numerical relativity simulations. In particular, we characterise the waveform differences between
    binary boson-star and black-hole systems across (i) the \textit{early inspiral}, by matching our waveforms to post-Newtonian expressions, (ii) \textit{merger}, and (iii) \textit{late ringdown}, by extracting the quasi-normal mode frequencies of the remnants. We find that boson-star binaries exhibit the largest deviations from comparable binary black-hole systems during the late inspiral and merger phases.
    Remarkably, for a subset of these equal-mass boson-star binaries (with certain phase offsets in the scalar-field profiles) we identify the excitation of subdominant odd $m$-multipoles in the gravitational-wave emission,
    absent in equal-mass nonspinning black-hole binaries. Despite differences in the phenomenology of binary boson-star and black-hole signals, 
    injections of some boson-star signals into detector noise exhibit degeneracy with current waveform approximants. 
    Building on these results, we demonstrate how inspiral-merger-ringdown consistency tests can overcome these degeneracies. 
\end{abstract}

\maketitle

\section{Introduction}

The advent of gravitational-wave (GW) astronomy has opened up exciting opportunities to test theories of gravity and explore the nature of compact objects inhabiting our universe. The latest LIGO–Virgo–KAGRA (LVK) catalogue~\cite{LIGOScientific:2025slb} already indicates that future observing runs will deliver a large number of loud detections, thereby allowing us to further constrain our theoretical models. Yet a fundamental question concerning tests of general relativity (GR) and compact objects remains: \textit{how can we prepare for genuinely unexpected discoveries}? One step forward in addressing this challenge is to assess the sensitivity of currently proposed GW tests~\cite{LIGOScientific:2021sio,LIGOScientific:2026qni,LIGOScientific:2026fcf,LIGOScientific:2026wpt,Johnson-Mcdaniel:2018cdu,Chia:2023tle,Krishnendu:2017shb,Krishnendu:2018nqa,Pompili:2025cdc,Johnson-McDaniel:2021yge} to common signatures or deviations arising from \textit{non-conventional} systems -- namely, sources other than vacuum
binary black holes or binary neutron stars within GR. To make this question tractable it is instructive
to focus on a specific, well-defined model. 

Motivated by this question, in this work we focus on a specific class of exotic compact objects -- boson stars (BSs)~\cite{Kaup:1968zz, PhysRevD.35.3658,Kleihaus:2005me}. First introduced in the late 1960s, BSs are self-gravitating configurations of ultralight scalar fields that have since gained popularity as \textit{proxies} to study compact objects with no horizon. The GW signatures of BSs have been investigated across a broad range of dynamical scenarios, including head-on collisions~\cite{Palenzuela:2006wp, Bezares:2017mzk, Helfer:2021brt, Evstafyeva:2022bpr, Ge:2024itl}, inspirals~\cite{Bezares:2022obu,Siemonsen:2023hko,Siemonsen:2023age, Evstafyeva:2024qvp} and scattering \cite{Damour:2025oys}, revealing a rich phenomenology (see e.g.~Refs.~\cite{Liebling:2012fv,Bezares:2024btu} for a review). In Ref.~\cite{Evstafyeva:2024qvp} we presented a set of waveforms based on numerical relativity (NR) simulations of inspiralling equal-mass non-spinning BS binaries with different compactness. Using these waveforms, we investigated
the response of GW analysis when such exotic signals are present in the data but interpreted under the binary black-hole (BBH) hypothesis. For less compact binaries, which exhibit distinct and characteristic GW features, significant deviations from black hole (BH) signals could be diagnosed straightforwardly through a residual test at large enough signal-to-noise ratio (SNR). However, we found in Ref.~\cite{Evstafyeva:2024qvp}
their GW signals to be fully degenerate with those of BBHs
(albeit with biased parameter estimation) for sufficiently compact configurations,
raising questions about the observational distinguishability of the two types of systems. 

In this paper our aim is to provide more information on the dynamics and GW emission of the binaries presented in Ref.~\cite{Evstafyeva:2024qvp}. In Section~\ref{sec:simulations} we present our NR simulations and discuss their accuracy in the context of GW observations. We next investigate the GW morphology of our numerically computed signals in more detail and compare them to the signatures of comparable BBH systems. In particular, (i) in Section~\ref{sec:inspiral} we compare how well the inspiral part of our simulations agrees with predictions from 3.5 post-Newtonian (PN) order theory, (ii) in Section~\ref{sec:merger} we discuss the relative excitation of higher harmonic modes around merger, and (iii) in Section~\ref{sec:ringdown} we analyse the late ringdown content of our waveforms.
After summarizing these features in
Sec.~\ref{sec:Intermission}, we investigate in Section~\ref{sec:imr} how they map onto inspiral–merger–ringdown consistency tests~\cite{Ghosh:2017gfp} through a campaign of injections of our NR waveforms into Gaussian noise. We thereby provide first steps towards assessing the ability of current GW analyses to capture or miss unexpected signatures of exotic compact objects.

In the following, we use natural units where $c=G=\hbar=1$, leaving
the scalar mass parameter $\mu$ as the only scale with dimension of
inverse length.

\section{NR simulations} \label{sec:simulations}

Boson stars are composed of a massive scalar field, minimally coupled to gravity via the action
\begin{equation} \label{eq:action}
S = \int \frac{\sqrt{-g}}{2} \left\{\frac{R}{8 \pi}- \left[g^{\mu \nu} \nabla_{\mu} \bar{\varphi} \nabla_{\nu} \varphi + V(\varphi) \right] \right \} \du ^4x.
\end{equation}
Here we take the scalar potential $V(\varphi)$ to be of solitonic form,
\begin{equation}
    V(\varphi)=\mu^2|\varphi|^2 \left(1-2\frac{|\varphi|^2}{\sigma_0^2} \right)^2,
\end{equation}
characterized by the scalar mass $\mu$ and the self-interaction parameter $\sigma_0$. In the remainder of this text, we set $\sigma_0=0.2$. Spherically symmetric BS models in equilibrium are
obtained by writing the scalar field in the form
\begin{equation}\label{eq:phi-anstaz}
    \varphi(t, r) = A(r)e^{\iu( \epsilon
\omega t + \delta \phi)},
\end{equation}
where $\omega$ is the constant BS frequency, $A(r)$ is the amplitude profile, $\epsilon = \pm 1$ and $\delta \phi$ is the initial phase offset. Individual BS solutions are computed using a shooting algorithm\footnote{Here we shoot for the BS frequency, $\omega$.} as for example described in Ref.~\cite{Helfer:2021brt}. 

\begin{table}[t]
\caption{Main properties of selected BS models in the $\sigma_0 = 0.2$ family of solutions. $A(0)$ denotes the central scalar-field amplitude (at $r=0$), $M_{\rm BS}$
the ADM mass, $\omega$
the BS frequency, $C\defeq \max \frac{m(r)}{r}$ the compactness defined in terms of the mass $m(r)$ enclosed inside areal radius $r$,
and
$k^{\rm{E}}_2$, $k^{\rm{B}}_2$ the $l=2$ polar (even) and axial (odd) tidal Love numbers of the progenitors, as defined by Eqs.~(C11) and (C16) of Ref.~\cite{Cardoso:2017cfl}, respectively. We recompute these tidal Love numbers using the methods outlined in Refs.~\cite{Cardoso:2017cfl,Sennett:2017etc,Hinderer:2007mb}.
}
\begin{tabular}{|c | c | c | c |c | c|} 
\hline
~~~$A(0)$~~~&~~~$\mu M_{\rm BS}$~~~&~~~$\omega/\mu$~~~&~~~$C$~~~&~~~$k_2^E$~~~ &~~~$k_2^B$\\ 
\hline
0.147 & 0.361 & 0.678 & 0.11 & 443 & -105 \\
0.155 & 0.439 & 0.605 & 0.14 & 141 & -42 \\
0.16 & 0.537 & 0.537 & 0.17 & 48 & -18 \\ 
0.17 & 0.713 & 0.439 & 0.22 & 7.4 & -4.7 \\ 
\hline
\end{tabular}
\label{tab:models} 
\end{table}

Binary initial data are constructed by superposing two equilibrium BS models with the \textit{constant volume element} technique, as described in Ref.~\cite{Helfer:2021brt}. We find this choice necessary for the faithful evolutions described below, as it (i) precludes premature gravitational collapse of compact stars in a binary, and (ii) significantly reduces spurious breathing oscillations of the scalar field. 
We utilise models of Table~\ref{tab:models} to construct such binary solutions.

Eccentricity is reduced iteratively using Eq.~(17) of Ref.~\cite{Mroue:2010re} and estimated to be $e \sim 0.002-0.005$. The initial BS configurations and their properties determine the type of the remnant formed post-merger. 
In our simulations of inspiralling equal-mass quasi-circular configurations, we empirically find \textit{fluffy} (less compact) BSs with $A(0) \leq 0.155$ in a binary to form BS remnants, whilst for $A(0) \geq 0.16$ we form a BH post-merger. 
Henceforth we focus on two representative families of binaries: \textit{compact} binaries with $A(0)=0.17$ which form a BH remnant post-merger, and \textit{fluffy} binaries with $A(0)=0.147$ forming a BS remnant. The stars with $A(0) = 0.17$ are located on the stable branch close to the maximum mass of the $\sigma_0 = 0.2$ potential, which we estimate to be around $A_{\rm max}(0) \sim 0.174$.

The scalar field ansatz \eqref{eq:phi-anstaz} carries many parameters that we are free to choose. To simplify our notation we will shorten the names of our simulations using the notation \texttt{Axx-dxx}, where \texttt{Axx} is the placeholder for the central amplitude and \texttt{dxx} the initial separation in units of the binary's rest mass $M\defeq 2M_{\rm BS}$. 
We will append \texttt{pxxx} (in degrees), whenever one of the stars in the binary has a non-zero phase offset, and \texttt{e1} when one of the stars has a frequency with the opposite sign, i.e.~$\epsilon=-1$, henceforth referred to as \textit{antiboson} (\textit{anti-BS}) binary. Further, we will refer to binaries with $\delta \phi = \pi$ as \textit{anti-phase} and binaries with generic $\delta \phi ~\mathrm{mod}~ \pi \ne 0$ as \textit{dephased}. 

\begin{table*}[t!]
\caption{
Summary of the BS  binaries evolved in the centre-of-mass
frame with initial boost velocity $v_{x, \rm{ini}}$ in the
$x$-direction, impact parameter $b$ in the $y$-direction and
offset
$d$ in the $x$-direction. The dephased ($\delta \phi ~\mathrm{mod}~ \pi \ne 0, \epsilon = 1$), anti-phase ($\delta \phi = \pi, \epsilon = 1$) and
anti-BS ($\delta \phi = 0, \epsilon = -1$) binaries are labelled using suffixes \texttt{p090} (for $\delta \phi=\pi/2$) and \texttt{p120} (for $\delta\phi=2\pi/3$),
\texttt{p180} and
\texttt{e1}, respectively. $E_{l=2}$ is the GW energy contained in the $l=2$
modes.
For comparison, a nonspinning, equal-mass black-hole binary  emits $\sim 5\%$ in GW energy \cite{Scheel:2008rj}. For binaries that form a BH remnant, we compute the final mass $M_{\rm fin}$ and final dimensionless spin $j_{\rm fin}$ of the BH using energy balance arguments. Where available we also report the Christodoulou mass $M_{\rm Chr}$ and dimensionless spin $j_{\rm AH}$ obtained from the apparent horizon finder data. With an asterisk we mark the simulation not previously reported in Ref.~\cite{Evstafyeva:2024qvp}.
}
\begin{tabular}{|c | c | c | c| c| c| c | c | c | c | c | c | c |} 
\hline
~~~Simulation~~~&~~~$v_{x, \rm{ini}}$~~~&~~~$b/M$~~~&~~~$d/M$~~~&~~~$E_{l=2}/M$~~~&~~~ $M_{\rm Chr}/M$~~~&~~~$j_{\rm AH}$~~~&~~~$M_{\rm fin}/M$~~~&~~~$j_{\rm fin}$~~~\\ 
\hline
\texttt{A17-d12} & 0.1671 & 12.283 & 0.2243 & 0.0333 & 0.9582 & 0.6941 & 0.9654 & 0.6998 \\ 
\hline
\texttt{A17-d12-p120}$^{*}$ & 0.1671 & 12.283 & 0.2243 & 0.0342 & -- & -- & 0.9650 & 0.6892 \\
\hline
\texttt{A17-d12-p180} & 0.1671 & 12.283 & 0.2243 & 0.0353 & -- & -- & 0.9638 & 0.6730 \\
\hline
\texttt{A17-d12-e1} & 0.1671 & 12.283 & 0.2243 & 0.0299 & -- & -- & 0.9691 & 0.6978 \\
\hline
\texttt{A17-d14} & 0.1533 & 14.017 & 0.2102 & 0.0346 & -- & -- & 0.9627 & 0.6958 \\
\hline
\texttt{A17-d15} & 0.14625 & 15.087 & 0.2033 & 0.0367 & 0.9506 & 0.7029 & 0.9709 & 0.7012 \\
\hline
\texttt{A17-d15-p090} & 0.14625 & 15.087 & 0.2033 & 0.0377 & 0.9540 & 0.6831 & 0.9701 & 0.6769 \\
\hline
\texttt{A17-d15-p180} & 0.14625 & 15.087 & 0.2033 & 0.0387 & 0.9539 & 0.6728 & 0.9689 & 0.6689 \\
\hline
\texttt{A17-d15-e1} & 0.14625 & 15.087 & 0.2033 & 0.0352 & 0.9533 & 0.7003 & 0.9720 & 0.7011 \\
\hline
\texttt{A147-d17} & 0.1389 & 16.639 & 0.0693 & 0.0589 & -- & -- & -- & -- \\
\hline 
\texttt{A147-d19} & 0.1256 & 19.412 & 0.1109 & 0.0691 & -- & -- & -- & -- \\ 
\hline
\end{tabular}
\label{tab:waveforms} 
\end{table*}

In Table~\ref{tab:waveforms} we summarise the set of BS binaries studied in the remainder of the text, including the radiated energy contained in the dominant $l=2$ multipoles of the Newman-Penrose scalar $\Psi_4$,
\begin{equation}
    \frac{\du E_{l=2}}{\du t} = \lim_{r \to \infty} \frac{1}{16 \pi} \sum_{l=2, m} \left| \int_{t'} r \Psi_{4,lm} \du t'\right|^2.
\end{equation}
Finally, for binaries that form a BH, we calculate the final mass and final spin of the BH using (i) energy balance arguments (cf.~e.g.~Eqs.~(19)-(20) of Ref.~\cite{Radia:2021hjs}) and (ii) apparent horizon data. Overall, the two measures agree within $\sim 2 \%$ in the final mass and $\sim 1 \%$ in the final dimensionless spin. 

\subsection{Computational infrastructure}

Our boson-star binaries (BBSs) are evolved with two independent codes: {\sc exozvezda} and {\sc lean}. Let us first summarise the main differences between the codes.

\begin{enumerate}[label=(\roman*)]
\item {\sc exozvezda} is an open-source code~\cite{exozvezda}, built as an extension of {\sc grchombo} \cite{Andrade2021,Radia:2021smk,Clough:2015sqa} and developed to evolve exotic compact objects. The adaptive-mesh-refinement is provided by the {\sc chombo} libraries~\cite{chombo}. For the compact \texttt{A17} binaries, the grid is refined by comparing gradients of the scalar field and the conformal factor with specified threshold values, while we use the \textit{moving boxes} approach detailed in Section~4.1.1 of Ref.~\cite{Radia:2021smk} for fluffy \texttt{A147} configurations. Empirically, we find the moving boxes approach to particularly help with the accuracy of evolutions of less compact binaries that form a BS remnant. Resolutions of $\mu \Delta x = 1/40$ on the finest level are employed for both binary types and GWs are extracted at $\mu R_{\rm ext} = 140$. 

\item 
The {\sc lean} code \cite{Sperhake:2006cy} is based on
the {\sc cactus} computational toolkit \cite{Allen:1999} and evolves
$3D$ domains with a \textit{box-in-a-box style} mesh refinement provided by {\sc carpet} \cite{Schnetter:2003rb}
using a resolution $\mu\Delta x$ in the range $1/32$ to $1/48$ on the
innermost level. Apparent horizons are computed with \textsc{ahfinderdirect} \cite{Thornburg:1995cp,Thornburg:2003sf}, and GWs are extracted at $\mu R_{\rm ex}=240$.
\end{enumerate}

In other aspects, both codes have identical set-ups. In particular, we use a cubic computational domain of length $\mu L = 1024$ with 8 refinement levels and bitant symmetry. At the boundaries, (radiative) Sommerfeld conditions are imposed. Further, both codes evolve the Einstein-Klein-Gordon system of equations using the covariant and conformal Z4 (CCZ4) formulation and employ fourth-order spatial differencing with a fourth-order Runge-Kutta integration in time. The CCZ4 damping terms are set to 
$\alpha \kappa_1 = 0.1 \mu$, 
$\kappa_2 = 0$ and $\kappa_3 = 1$. Lapse and shift are evolved using the moving puncture gauge, as given by Eq.~(18) of Ref.~\cite{Radia:2021smk}. We have also performed a few simulations using the Baumgarte-Shapiro-Shibata-Nakamura-Oohara-Kojima (BSSN) formulation \cite{Shibata:1995we,Baumgarte:1998te}. However, despite the advantages of using the constant volume technique for initial data,
BSSN runs produced order of magnitude larger oscillations in the maximal scalar field amplitude $A_{\rm max}(t)$
compared to CCZ4. This is likely a result of better constraint preservation in CCZ4, which could also explain why we did not succeed in accurately circularising inspirals in BSSN.

\subsection{Mismatch} \label{sec:mismatch}
Having presented the set of numerical waveforms we study here, we now complement the convergence analysis of the amplitude and phase of GW signals from \texttt{A17} and \texttt{A147} binaries presented in the Supplemental Material of Ref.~\cite{Evstafyeva:2024qvp}. Specifically, we aim to understand the accuracy of our waveforms
in the scope of GW observations. In particular, our expectation is that the numerical error will vary across different parts of the waveform which, in LVK injections, translates into a frequency-dependent error controlled by the binary's total mass. When mapped onto the sensitivity curves of GW detectors, these variations can lead to different accuracy estimates. We therefore quantify the quality of our data in the context of GW observations by calculating the waveform \textit{mismatch}.

The mismatch is defined as 1 minus the overlap between two normalized waveforms, $h_1$ and $h_2$, maximised over relative time $t$ and GW phase shifts $\Phi$,
\begin{equation} \label{eq:mismatch}
    \mathcal{M} = 1 - {\rm{max}}_{\Phi, t} \frac{(h_1, h_2)}{\sqrt{(h_1, h_1)(h_2, h_2)}}.
\end{equation}
Here $(h_1, h_2)$ denotes the noise-weighted inner product,
\begin{equation}
    (h_1, h_2) = 4 \mathrm{Re} \left\{\int_{f_{\rm{low}}
    }^{f_{
    \rm{high}
    }} \frac{\tilde{h}_1(f) \tilde{h}_2(f)}{S_{
    \rm{n}}(f)} \du f \right\},
\end{equation}
where $S_{\rm{n}}(f)$ is the one-sided power spectral-density (PSD) of the noise, which we take to be that of the LIGO Hanford detector's O4 design sensitivity~\cite{KAGRA:2013rdx}. Furthermore, $f_{\rm low}$ is defined as the larger of the waveform's starting frequency and 20\,Hz, while $f_{\rm high}=2048 \, \rm{Hz}$.

Our mismatch calculation focuses on the comparison between (i) waveforms at different resolutions in the {\sc exozvezda} and {\sc lean} codes, separately -- henceforth referred to as \textit{resolution} comparison -- and (ii) {\sc exozvezda} and {\sc lean} waveforms at equal resolutions -- henceforth referred to as \textit{code} comparison. 
We note that the latter comparison is affected by small differences in the initial data and radii for waveform extraction used in the codes\footnote{For instance, as can be seen from Table~\ref{tab:waveforms}, 
the two codes' evolutions of otherwise identical binaries
start at different separations and accordingly adjusted initial velocities.}. We therefore expect larger mismatches in the code comparison. Before performing the Fourier transform of the strain to calculate the mismatch \eqref{eq:mismatch}, we remove junk radiation and taper the start of our waveforms.
We then scale the waveforms to binary masses in the range $M_{\rm{tot}} \in [10, 150]M_{\odot}$ and apply phase and time shifts to $h_1$, until the overlap (mismatch) is maximised (minimised).

\begin{figure}[t!]
    \includegraphics[width=0.9\linewidth]{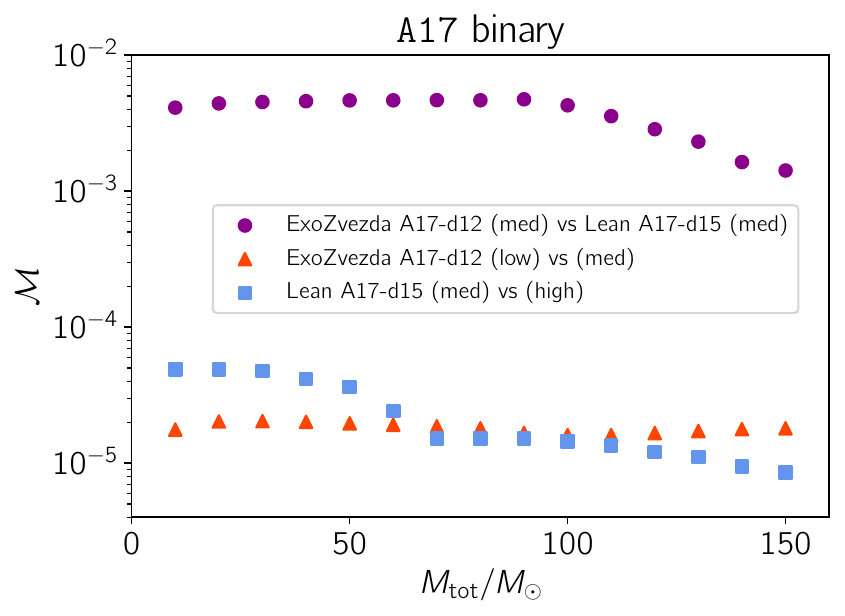} \\
     \includegraphics[width=0.9\linewidth]{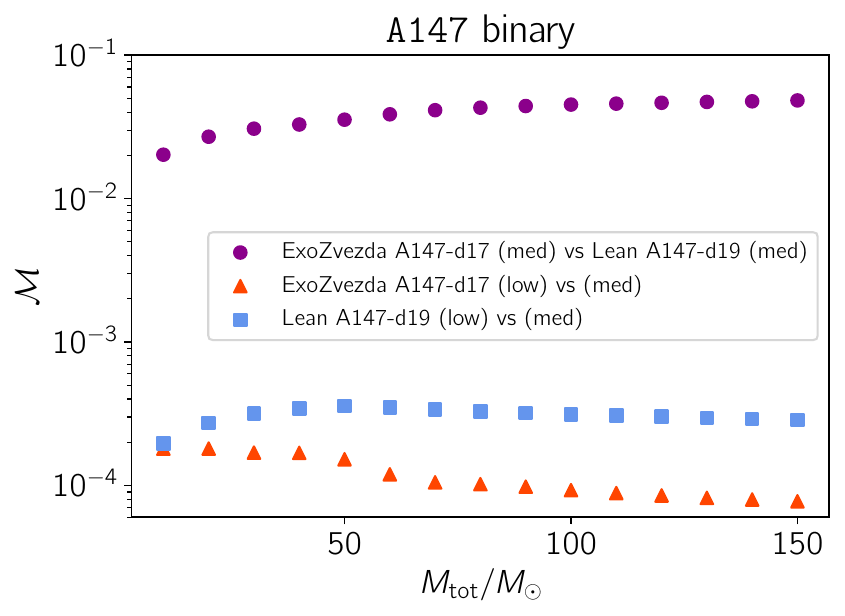}
    \caption{
    \textit{Top}: Mismatch as defined in Eq.~\eqref{eq:mismatch} for the \texttt{A17} binary as a function of total binary mass $M_{\rm{tot}}/M_{\odot} \in [10,150]$. High, medium and low resolutions correspond to $\mu \Delta x = 1/48$, $\mu \Delta x = 1/40$, and $\mu \Delta x = 1/32$ on the innermost levels, respectively. Note that the {\sc lean} and {\sc exozvezda}
    codes use different initial separations and velocities for
    otherwise identical binaries and extract waveforms at different radii, hence incurring a larger mismatch in the code comparison. \textit{Bottom}: The same as the upper panel, but for the \texttt{A147} binary.
    }
    \label{fig:mismatch-A17}
\end{figure}

Focusing on the (22)-mode only, we summarize in Figure~\ref{fig:mismatch-A17} our mismatch findings for resolution and code comparison.  
Overall, comparison across the two codes results in mismatches of at most $\sim 5 \times 10^{-3}$ for the compact \texttt{A17} binary and $\sim 5 \times 10^{-2}$ for the fluffy \texttt{A147} binary. Comparison of the same configurations at different resolutions, on the other hand, produces smaller mismatches, as expected. The shorter compact binary \texttt{A17-d12} results in a mismatch of at most $ \mathcal{M}_{\rm max} \sim 2 \times 10^{-5}$ at low ($\mu \Delta x = 1/32$) and medium ($\mu \Delta x = 1/40$) resolutions, while the longer binary \texttt{A17-d15} at medium ($\mu \Delta x = 1/40$) and high ($\mu \Delta x = 1/48$) resolutions yields at most $\mathcal{M}_{\rm max} \sim 5 \times 10^{-5}$. For the \texttt{A147} family, we incur a larger mismatch of $\mathcal{O}(10^{-4})$ in the resolution comparison. This is not unexpected, as we measure a larger numerical error budget for binaries that form a BS remnant, in part due to the fact that these simulations cover more orbits \cite{Evstafyeva:2024qvp}.

With the set of different binary mismatches at hand, we can now make a projection for the maxmimum of the squared SNR, at which the two waveforms cannot be distinguished. For this purpose, we utilise the approximate criterion~\cite{Thompson:2025hhc,McWilliams:2010eq,Purrer:2019jcp}, 
\begin{equation} \label{eq:snr_criterion}
    \rho^2 (\mathcal{M}) \lesssim 
    \frac{D}{2\mathcal{M}},
\end{equation}
where $D$ is a constant prefactor, often defined as the number of intrinsic parameters whose measurement is affected by the waveform inaccuracies~\cite{Scheel:2025jct,Purrer:2019jcp}. We note that Eq.~\eqref{eq:snr_criterion} is approximate and assumes that $\rho^2 = (h_1, h_1) = (h_2, h_2)$~\cite{Toubiana:2024car}.
However, when this inequality is satisfied, it ensues that systematic errors from waveform
inaccuracies are smaller than 1-sigma (i.e.~standard deviation) statistical errors. 

In the scenario of synthetic injections of BBS signals into detector noise, at present only BBH (or inspiral BNS)
templates can be used in recovery, due to the lack of available alternatives. 
Therefore, for the most generic BBH system, we pick $D=8$, representing the mass ratio, total mass, and six spin components of the primary and the secondary. Focusing on the mismatches computed in Fig.~\ref{fig:mismatch-A17} at different resolutions (but within the same code) and utilising Eq.~\eqref{eq:snr_criterion} with $D=8$, we estimate an upper bound\footnote{For reference, the currently loudest GW event observed is
GW250114 with a network SNR in the range 77 to 80 \cite{LIGOScientific:2025rid}.} of $\rho(\mathcal{M_{\rm max}}) \sim 280$ for the \texttt{A17} binary and $\rho(\mathcal{M_{\rm max}}) \sim 115$ for \texttt{A147}. 

\section{Act I: Phenomenology of BBS binaries}

In this section we provide more details on the inspiral (Section~\ref{sec:inspiral}), merger (Section~\ref{sec:merger}) and late ringdown (Section~\ref{sec:ringdown}) characteristics of the binaries studied here.

\subsection{Early inspiral: comparison with PN results} \label{sec:inspiral}

\begin{figure*}[t!]
    \centering
    \includegraphics[width=\linewidth]{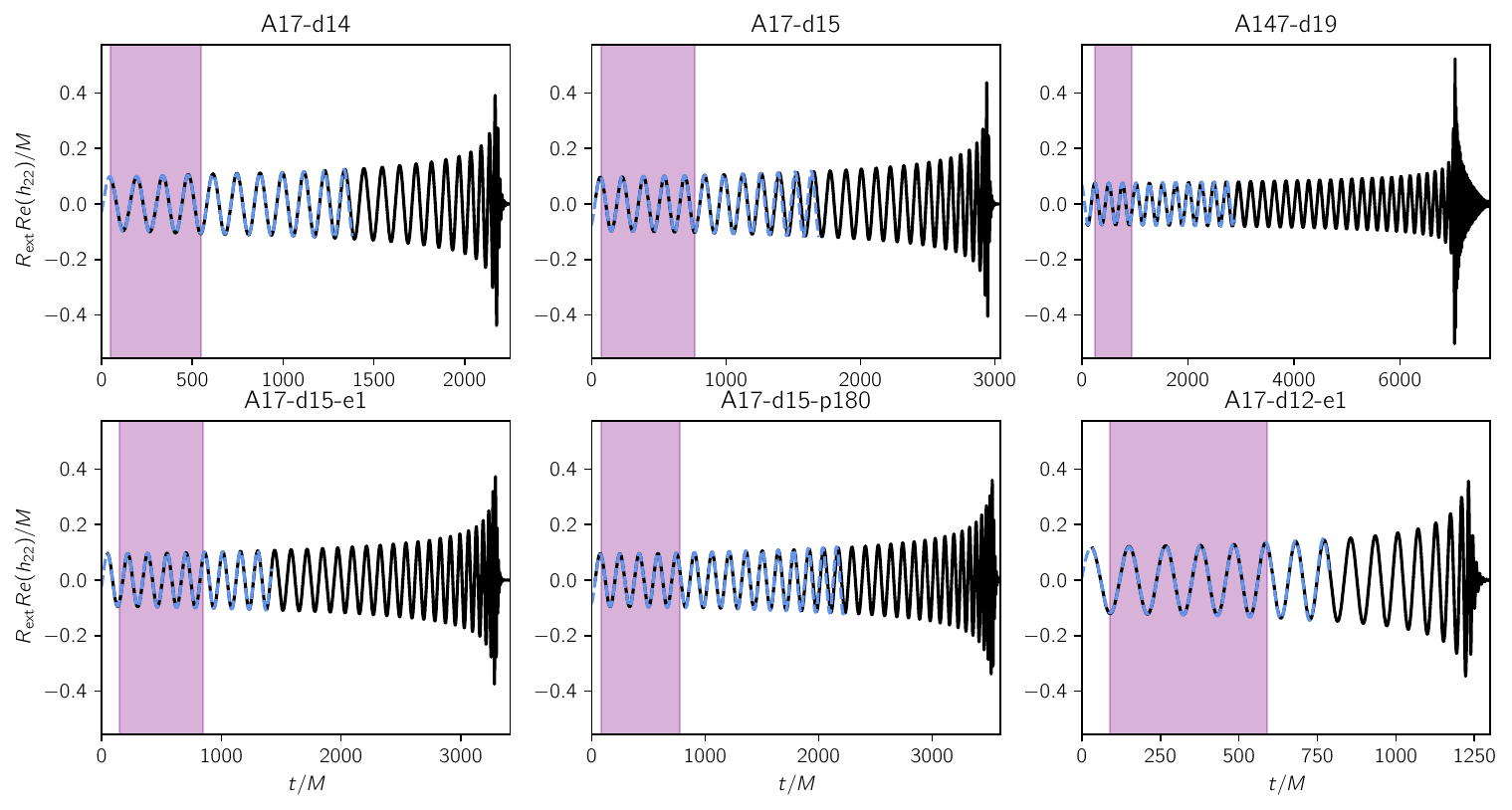}
    \caption{
    Comparison between 3.5PN (dashed light blue) and NR (black) binaries listed in Table~\ref{tab:waveforms}. We illustrate in this comparison the real part of the $(22)$-mode, however the results for the imaginary parts of the waveforms look (up to their usual $\pi/2$ phase shift) identical. In light purple we show the matching window.
    } 
    \label{fig:pn_match}
\end{figure*}

We start by comparing our suite of NR waveforms with 3.5PN estimates for non-spinning quasi-circular BBHs, following expressions provided in Ref.~\cite{Blanchet:2013haa}. We leave a more comprehensive comparison with higher PN effects (including e.g.~tidal effects, as quantified in Table~\ref{tab:models}) and the effective-one-body formalism (EOB)~\cite{Buonanno:1998gg} for future work. The PN expansion is given in powers of the dimensionless  variable $x = (M \Omega)^{2/3}$, where $\Omega$ denotes the orbital frequency, and terms $\propto x^n$ correspond to $n/2$-th PN order. Here we consider the 3.5PN expansion up to and including $x^{7}$ terms. 

In our comparison, we construct the corresponding PN $(lm)$ modes as projections of the GW polarizations onto the basis of spin-weighted spherical harmonics. For the dominant $(22)$-mode, we thus obtain
\begin{equation}
    h^{22}(x, \Phi) = 2 \tilde{\mu} x M \sqrt{\frac{16 \pi}{5}} H^{22}e^{-2\Phi}, 
\end{equation}
where $\tilde{\mu} = (m_1 m_2)/M$ is the reduced mass,
$H^{22}$ is the amplitude and $\Phi$ is the orbital phase.  To reconstruct $H^{22}$ and $\Phi$ we utilise Eq.~(491) of Ref.~\cite{Blanchet:2013haa}. In order to evaluate $h^{22}(x, \Phi)$ as a function of time, however, we also need to evolve $x$ and $\Phi$ forward in time by numerically integrating their corresponding ordinary differential equations (ODEs). This is done by recalling the energy balance 
\begin{equation}
    \frac{\du E}{\du t} = \mathcal{F},
\end{equation}
where $E$ is the binding energy as in Eq.~(377) of Ref.~\cite{Blanchet:2013haa} and $\mathcal{F}$ is the GW energy flux given by their Eq.~(483). Then, the required ODE for $x$ immediately follows from chain rule,
\begin{equation}
    \frac{\du x}{\du t} = -\mathcal{F} \left(\frac{\du E}{\du x}\right)^{-1}, 
\end{equation}
which is supplemented by the equation for the orbital phase $\Phi$
\begin{equation}
    \frac{\du \Phi}{\du t} = \Omega = \frac{x^{3/2}}{M}. 
\end{equation}

To hybridize our PN and NR waveforms, we need to choose a matching window; our specific choice is guided as follows. Since PN estimates are generally less accurate towards late inspiral and merger, we start the matching as early as possible (e.g.~after removal of junk radiation). The optimal length of our matching window, $L_{\rm opt}$, is chosen by inspecting the absolute differences between the PN and NR GW phases, $\Delta \phi_{\rm abs} \defeq |\phi^{\rm NR} - \phi^{\rm PN}|$. Employing a threshold $\Delta \phi_{\rm abs} \lesssim 0.01$, we obtain $L_{\rm opt} \sim 700\,M$ and $L_{\rm opt} \sim 500\,M$ 
for long and short runs, respectively.
Defining the relative amplitude variation\footnote{As commonly done in hybridizing waveforms, we scale the PN amplitude by a constant factor to optimally match it to the NR amplitude inside the matching window; see e.g.~Refs.~\cite{Sperhake:2011zz, Pan:2007nw, Ajith:2007kx, Ajith:2007qp}.} $\Delta A_{\rm rel} \defeq |A_{\rm PN} - A_{\rm NR}|/ A_{\rm NR}$, we find $\Delta A_{\rm rel} \lesssim 8 \times 10^{-3}$ 
inside the matching window for longer waveforms (e.g.~\texttt{A17-d15}) and $\Delta A_{\rm rel} \lesssim 5 \times 10^{-2}$ for shorter waveforms (e.g.~\texttt{A17-d12-e1}). 
In Fig.~\ref{fig:pn_match}, we graphically illustrate the 3.5PN hybridization,
including the near-perfect overlap of
the PN and NR signals inside the matching window. 

\subsection{Merger} \label{sec:merger}

The merger part of BS-binary evolutions can presently only be studied using numerical relativity techniques. The GW emission of these binaries around merger depends on two main features: (i) the character and properties of the merger remnant (BS or BH), and (ii) the nature of short-range scalar interactions encoded in the initial $\delta \phi$ and $\epsilon$ parameters. In the \texttt{A17} family of runs, these interactions manifest themselves in a
different shape of the `chirp' during late inspiral and merger (see e.g.~Fig.~S3 of Ref.~\cite{Evstafyeva:2024qvp}). In particular, in-phase binaries have a steeper chirp than binaries with non-zero phase off-sets, while the anti-BS binaries most closely reproduce the waveforms and corresponding chirp-mass parameters from equal-mass non-spinning BBH systems. These distinct binary characteristics will become important for interpreting results of parameter estimation in our full Bayesian inference setting in Section~\ref{sec:imr}.

Even though the quadrupole radiation from \texttt{A17} systems is barely distinguishable by eye from BBH waveforms, BS binaries with a dephasing $\delta \phi ~\mathrm{mod}~ \pi \ne 0$
differ qualitatively from their BH counterparts
through the presence of odd-$m$ content in their GW emission near merger 
(see also Ref.~\cite{Sanchis-Gual:2022mkk} for a similar effect in equal-mass head-on collisions of Proca stars). This odd $m$-content can be qualitatively understood as a consequence of the breaking of the exchange symmetry near the merger,
corresponding to a rotation by $\pi$ around the $z$-axis,~i.e $R_z(\pi) : (x, y, z) \to (-x, -y, z)$. For the trivial case $\delta \phi = 0$, the system and the associated stress–energy tensor are invariant under $R_z(\pi)$ and hence no odd-$m$ multipoles are present. For $\delta \phi = \pi$, the complex scalar field profile $\varphi$ 
changes sign under $R_z(\pi)$, but, the stress-energy tensor, which depends on $\varphi$ quadratically, remains invariant and the spacetime preserves the 
$\pi$-rotation symmetry -- odd-$m$ multipoles are therefore absent. In contrast, for generic relative phases $\delta \phi ~\mathrm{mod}~ \pi \ne 0$,
neither the scalar-field profile nor the stress–energy tensor remain invariant under $R_z(\pi)$, allowing for the excitation of odd-$m$ modes. It is important to note that this effect is negligible until the late inspiral and merger because the interaction (overlap) terms in the stress–energy tensor become dynamically important only
when the stars are close to each other.

\begin{figure}
    \centering
    \includegraphics[width=0.9\linewidth]{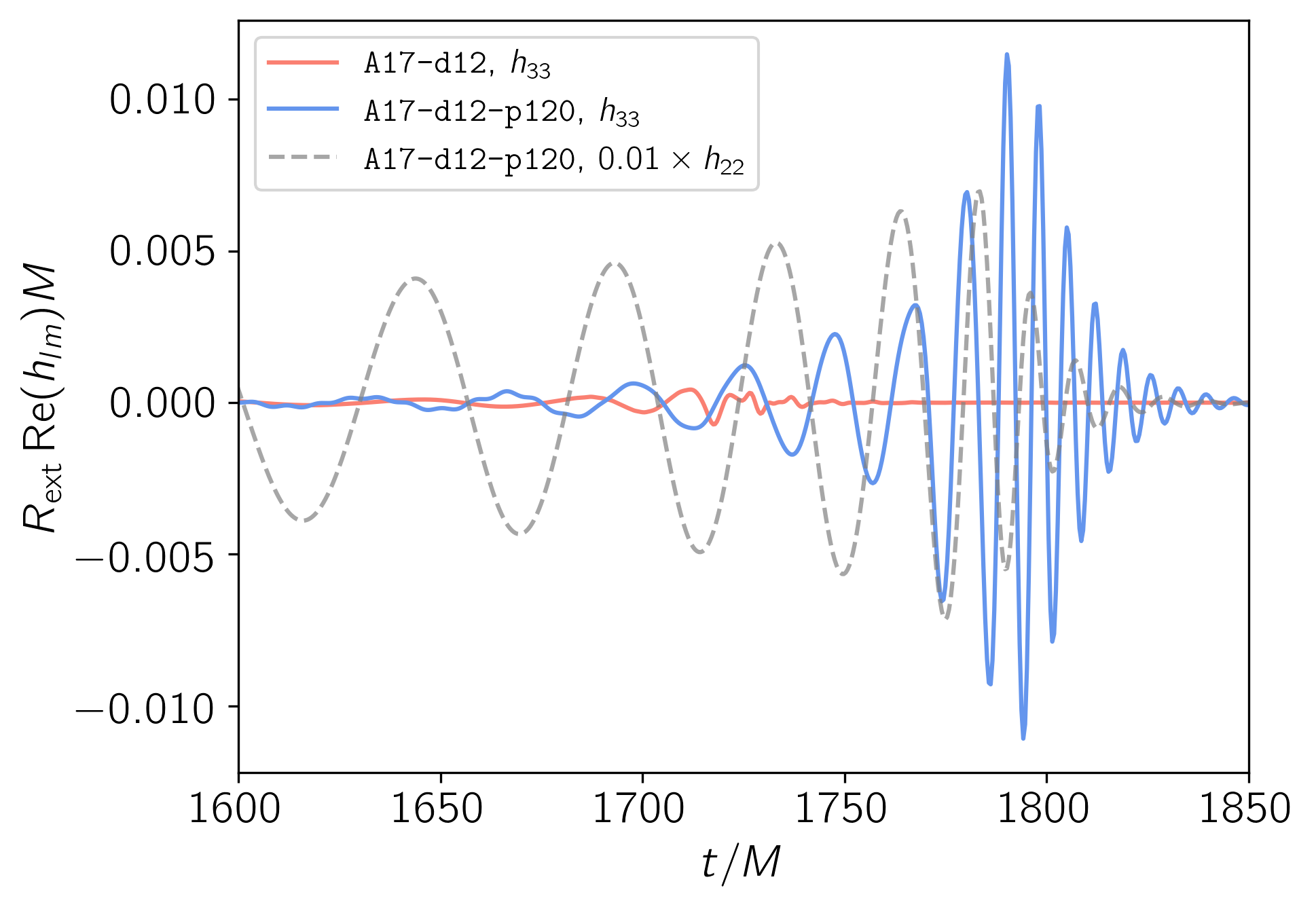}
    \caption{The real part of the (33)-mode of the \texttt{A17-d12} and \texttt{A17-d12-p120} binaries. While the (33)-mode is absent for the in-phase binary, it is present for $\delta \phi = 2\pi/3$ with the strain amplitude approximately two orders of magnitude smaller around merger when compared to its quadrupolar counterpart (shown in grey). We note that the (33)-mode of \texttt{A17-d12-p120} binary peaks $\sim 15M$ later than the (22)-mode, perhaps due to suppression in the scalar-field overlap at earlier times.
    }
    \label{fig:32-modes}
\end{figure}

In Figure~\ref{fig:32-modes} we illustrate the (33)-mode for the $\delta \phi = 2\pi/3$ binary\footnote{We have verified that the mode is physical, albeit small, across different resolutions.}. 
For binaries with phase offset 0 or $\pi$, this mode is absent.
The (33) contribution 
is small, however, approximately two 
orders of magnitude below the (22)-mode. In general, we can summarize the presence of
subdominant
multipoles as follows: the largest contribution comes from $l=4$, in particular the (44) mode, followed by the (32) mode as well as the (33) and (21) modes (if present), respectively. 
This behaviour reveals that excitation of odd $m$-modes can be a generic, albeit small, effect for BS binary systems with non-trivial initial phase offsets; cf.~Refs.~\cite{Puecher:2022sfm,Gupta:2025paz} for tests of GR involving higher-order GW modes. 

\subsection{Ringdown} \label{sec:ringdown}

We complete this section with a discussion of the ringdown and, in particular, its frequency content. Quasinormal modes (QNMs) of BSs have been first computed by Yoshida et al~\cite{Yoshida:1994xi} in the polar (even) sector and later have been extended by Macedo et al~\cite{Macedo:2013jja} to the axial (odd) case\footnote{We note that, contrary to the case of Schwarzschild BHs, the polar and axial QNMs of nonspinning BSs are not isospectral.}. An analogy can be drawn between the mode content of BSs and ordinary (fluid) stars, with the BS's scalar field playing the role of an anisotropic fluid~\cite{Macedo:2013jja}. The complex eigenfrequencies of BSs, however, are qualitatively different from those of fluid stars: the main differences stem from the absence of the proper surface in BS spacetimes which gives scalar perturbations the freedom to propagate to infinity. Whether we excite such characteristic frequencies of course 
depends on the type of the remnant we form. In this section, our aim is to extract QNMs at late times directly from the NR data.  

We approximate the late time behaviour of our numerical GW signals with a linear combination of QNMs with complex frequencies $\tilde {\omega} = \omega_{r,k} - \iu\omega_{i, k} = \omega_{r,k} - \iu/\tau_{k}$, where the $\tau_k$ denote the \textit{damping times},
\begin{equation} \label{eq:ringdown}
    h(t) = \sum_{k \in K} \mathcal{A}_k e^{-\omega_{i,k}t} e^{-\iu \omega_{r,k}t}.
\end{equation}
Here the $\mathcal{A}_k$ denote the complex amplitudes and 
$K$ the set of $(lmn)$ modes used in the fit, where $n$ corresponds to the overtone number. 
The cardinality of the set of modes will be denoted by $n(K)$. We focus on extracting the ringdown frequencies only from the fundamental $(lm) = (22)$ mode{\footnote{Of course we also have a mirror mode $l = 2, m= -2$. However, by symmetry, its ringdown frequency content is identical to the $l = 2, m= 2$ mode up to the sign difference in the real frequencies.} of our NR data, i.e.~we fit Eq.~\eqref{eq:ringdown} directly to $h^{22}_{\rm NR}$; henceforth we drop the `22' superscript and write $h_{\rm NR}$. We define the merger time\footnote{We find this measure of the merger to be consistent with the time when the two `surfaces' of the BSs come into close contact.} $t_{\rm merger}$ as the time at which the maximum of the strain amplitude, ${\rm max}(|h_{\rm NR}|)$, is attained.
The value of the ringdown starting time (i.e.~when we start to fit Eq.~\eqref{eq:ringdown}) is denoted by $t_0$. 

We perform the fitting procedure as outlined in Ref.~\cite{Cheung:2023vki}, using the ringdown fitting package \texttt{jaxqualin} as follows. We start with a \textit{free-mode} fit, where we vary the number of modes $n(K)\in[1,4]$. This allows us to identify all possible mode combinations present in the waveform. This type of fitting is performed across multiple starting times $t_0$ and therefore determines the evolution of QNM frequencies in time. After identifying all the possible modes for each $n(K)$, we fix their frequencies to the values determined in the free-mode fit,
and only fit for their real amplitudes $A_{k} \defeq |\mathcal{A}_k|$ and phases $\phi_{k} \defeq \rm{arg}(\mathcal{A}_k)$
across different starting times $t_0$.
We iteratively remove modes whose amplitudes and phases exhibit too much variation in time (designated as ``unstable'') and stop once we have identified all stable modes. 
Finally, we quantify the ``goodness'' of the fit using the fractional root mean square error~\cite{London:2014cma}
\begin{equation} \label{eq:error_rms}
    e_{\rm{rms}} = \left|\frac{\langle (h_{\rm{NR}} - h)^2\rangle}{\langle h_{\rm{NR}}^2\rangle }\right|^{1/2},
\end{equation}
where $\langle \cdot \rangle$ denotes the mean over the fitting window. 

Using the procedure outlined above, we are able to confidently identify only the fundamental (22) mode
in the NR data of our \texttt{A17} and \texttt{A147} binaries. We have tentative evidence that there is additionally a second mode present in the data, which also has a faster damping time (cf.~Appendix~\ref{app:two_mode_fits} for more details). However, we are not able to faithfully extract this mode due to its larger variations in the fitted frequencies across different starting times. Therefore, in Table~\ref{tab:ringdown} we present the results of a single-mode, $n(K)=1$, fit only. The $e_{\rm rms}$ values
therein 
clearly indicate that the ringdown content of both types of signals, \texttt{A17} and \texttt{A147},
can be characterised reasonably well by the damped sinusoid of Eq.~\eqref{eq:ringdown}. All the frequencies reported in the table are measured at the fiducial start-time $t_0 = 10M$, approximately corresponding to the time of AH formation. 
There is undoubtedly a trade-off between starting the fit too early -- away from the linear regime -- or too late -- deep in the region, where we are dominated by the truncation error. We therefore quantify the relative differences (with respect to our fiducial $t_0 = 10M$)
in complex frequencies measured at $5M \leq t_0 \leq 20 M$ ($5M \leq t_0 \leq 40M$) for the \texttt{A17} (\texttt{A147}) types of binaries. 
For instance, for the \texttt{A17-d12} binary we find variations of $M \delta \omega, M \delta \tau \lesssim  5\%$. On the other hand, for the \texttt{A147-d17} binary we estimate $M \delta \omega \lesssim 0.2\%$ and $M \delta \tau \lesssim 15\%$.
In Fig.~\ref{fig:ringdown_plot} we additionally illustrate the ringdown fit at various starting times for these representative compact \texttt{A17-d12} and fluffy \texttt{A147-d17} binaries.

The different remnants of the \texttt{A17} and \texttt{A147} binaries leave distinct imprints on the emitted post-merger frequencies. In particular, the ringdown of \texttt{A147} binaries is characterised by much longer damping time-scales than that of the \texttt{A17} configurations. The QNMs we extract for \texttt{A147} are quantitatively similar to the polar modes of BSs reported in Table V of Ref.~\cite{Macedo:2013jja}. For binaries with a BH remnant, in contrast, we find that their ringdown damping times are much shorter (more `BH-like'). Comparing their QNMs to the predicted fundamental (22) QNM of a Kerr BH in Table~\ref{tab:ringdown}, we systematically find  larger deviations in the damping times. The largest discrepancy from the Kerr predictions occurs for zero phase-offset: \texttt{A17-d12}, \texttt{A17-d14}
and \texttt{A17-d15} in Table \ref{tab:ringdown} yield deviations up to 
2\% in the real frequency and 13\% in the damping time. 
A small discrepancy between our measured frequencies and those from Kerr predictions is not unexpected, of course, due to errors stemming from numerical simulations and fitting. However, deviations of $\mathcal{O}(5-10) \%$ in some cases are too large to be explained solely by these errors. We now comment why some deviations from the Kerr spectrum may be present in our data. 

Due to 
our definition of $t_{\rm merger}$, early ringdown starting times may partially overlap with a dynamical phase, when the perturbed BS remnant has not yet collapsed to a BH or the scalar field has not completely dissipated or fallen into it.
We recall, in this context, that our BS binaries result
in a BH remnant only for $A(0)\ge 0.16$. This threshold is higher
than that observed for BH formation in \textit{head-on} collisions
in Ref.~\cite{Ge:2024itl} and suggests that residual angular
momentum from inspirals can support hypermassive BSs after merger,
a phenomenon well known for neutron stars \cite{Duez:2004nf}.
\label{page:hypermassive} 
For example, for the \texttt{A17-d12} binary, we find the apparent horizon forming around $\sim 10M$ after $t_{\rm merger}$, and the maximum of the scalar field amplitude, computed across the grid, dropping below $10^{-2}$ only after $t-t_{\rm merger} \gtrsim 40M$.
Therefore, although we find Eq.~\eqref{eq:ringdown} to provide a reasonable empirical fit to the waveform, its fitted frequencies across different $t_0$ should be interpreted with caution, especially when obtained from a fitting window that overlaps
with the collapse in the post-merger. This undoubtedly warrants further investigations on the time window where the validity of Eq.~\eqref{eq:ringdown} and the `linear' regime are fully justified. 

The amount of scalar debris left over around the BH post-merger, albeit ubiquitously
small, varies with the initial phase off-set $\delta \phi$. In general, we find that compact binaries with $\delta \phi \neq 0$ radiate more angular momentum in GWs than in-phase systems, hence leaving less angular momentum to support the remaining scalar field against the newly formed BH's pull. Furthermore, dephased binaries have smaller final BH spins and, hence, larger horizon radii capable of devouring
more scalar matter. Both effects clearly favor the $\delta \phi \neq 0$ binaries leaving less scalar field around the BH, than in-phase ones\footnote{This is also in agreement with our numerical data tracking the maximum of the scalar field across the whole grid.}, hence resulting in smaller differences in the ringdown when compared to Kerr. 

\begin{figure}
    \centering
    \includegraphics[width=0.95\linewidth]{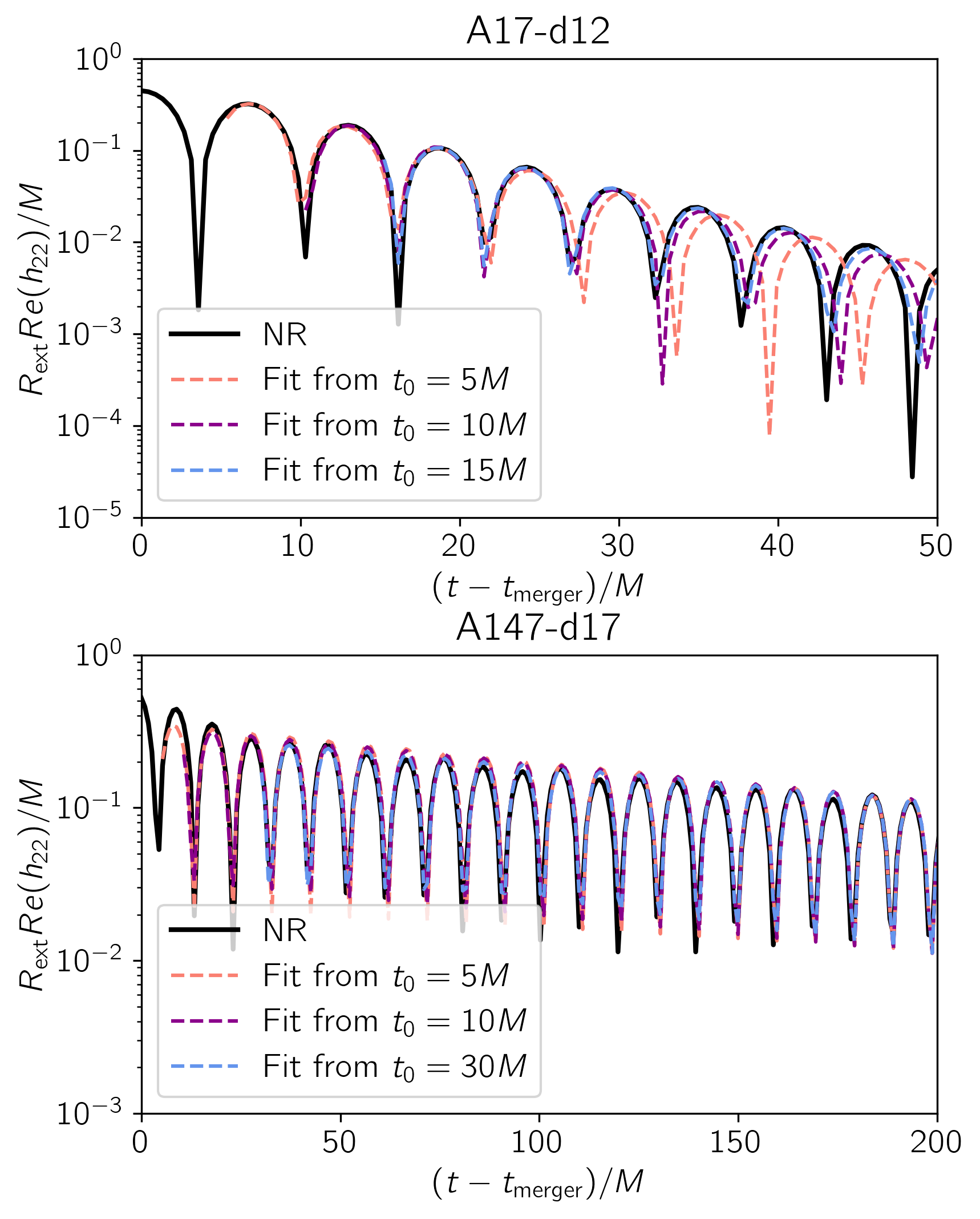}
    \caption{Examples of single-mode ringdown fits (as described in the main text) for the \texttt{A17-d12} (top panel) and \texttt{A147-d17} (bottom panel) binaries. We overplot the results of 
    the fits for the fiducial ringdown start-time $t_0 = 10M$ and other choices of $t_0$ for comparison. As can be clearly seen in the top panel, the fit for the compact binary significantly deteriorates for small $t_0$. For the fluffy configuration the dependence of the quality of the fit on the starting time is less evident by eye. However, by inspecting $e_{\rm rms}$ for different ringdown starting times we also find improvements in the fit for larger $t_0$. 
    }
    \label{fig:ringdown_plot}
\end{figure}

\begin{table*}
\caption{Numerical fits to the ringdown of our NR simulations as obtained with a one-mode fit, starting from $t_0/M = 10$ after the peak. For binaries forming a BH remnant we additionally list the predicted fundamental complex frequencies of a Kerr BH with the same final spin, as computed with BH perturbation theory~\cite{Berti:2009kk,Regge:1957td,Zerilli:1970se,Zerilli:1970wzz,Leaver:1985ax}. Here, $M\omega^{\rm Kerr}_{220}$ and $\tau^{\rm Kerr}_{220}/M$ denote the estimates computed using energy-balance arguments, whilst $M\omega^{\rm Kerr, AH}_{220}$ and $\tau^{\rm Kerr, AH}_{220}/M$ are determined using the apparent horizon data.
We also report the fractional root mean square error, $e_{\rm{rms}}$, defined in Eq.~\eqref{eq:error_rms}, to quantify the goodness of the fit.
}
\begin{tabular}{|c | c | c | c | c | c | c | c |} 
\hline
~~~Simulation~~~&~~~$M\omega_{r, 1}$~~~&~~~$\tau_{1}/M$~~~&~~~$M\omega^{\rm Kerr}_{r,220}$~~~&~~~$\tau^{\rm Kerr}_{220}/M$~~~&~~~$M\omega^{\rm Kerr, AH}_{r,220}$~~~&~~~$\tau^{\rm Kerr, AH}_{220}/M$~~~& ~~~$e_{\rm{rms}}$~~~\\ 
\hline
\texttt{A17-d12} & 0.564 & 10.36 & 0.552 & 11.95 & 0.553 & 11.83 & 0.099 \\
\texttt{A17-d12-p120} & 0.550 & 12.19 & 0.547 & 11.89 & -- & -- & 0.146 \\
\texttt{A17-d12-p180} & 0.552 & 11.93 & 0.541 & 11.79 & -- & -- & 0.122 \\
\texttt{A17-d12-e1} & 0.538 & 11.68 & 0.549 & 12.0 & -- & -- & 0.194 \\
\hline
\texttt{A17-d14} & 0.557 & 10.39 & 0.553 & 11.92  & -- & -- & 0.099 \\
\hline
\texttt{A17-d15} & 0.549 & 10.58 & 0.549 & 12.02 & 0.562 & 11.78 & 0.120 \\
\texttt{A17-d15-e1} & 0.524 & 13.13 & 0.548 & 12.03 & 0.559 & 11.80  & 0.261 \\
\texttt{A17-d15-p090} & 0.531 & 11.86 & 0.539 & 11.71 & 0.551 & 11.72 & 0.193 \\
\texttt{A17-d15-p180} & 0.546 & 12.31 & 0.536 & 11.83 & 0.546 & 11.67 & 0.108 \\
\hline
\texttt{A147-d17} & 0.322 & 175 & n/a & n/a & n/a & n/a & 0.262 \\
\texttt{A147-d19} & 0.318 & 177 & n/a & n/a & n/a & n/a & 0.203 \\
\hline
\end{tabular}
\label{tab:ringdown} 
\end{table*} 

\section{Intermission}
\label{sec:Intermission}
Having analysed in more detail the structure of our simulated GW signals from inspiralling BBS mergers, it is clear that the fluffy \texttt{A147} binaries are strikingly different from BBH systems, whereas the compact binaries \texttt{A17}, in so far as they differ from BH binaries, do so most significantly in the late inspiral and merger. 
However, as was shown in Ref.~\cite{Evstafyeva:2024qvp}, injections of GW signals from these compact \texttt{A17} binaries into the detector noise can easily be mistaken for BBHs due to the underlying \textit{systematic} signal degeneracies. For example, an in-phase BS binary exhibits strong degeneracy with unequal-mass BBHs with anti-aligned spins~\cite{Evstafyeva:2024qvp}. Therefore, even though we know \textit{apriori} that the compact BBS is not a BBH, in a realistic GW setting it may be challenging to differentiate the two systems on a single-event basis without additional checks. 

Since we expect deviations from BBH dynamics to arise most prominently in the late inspiral and merger of BSs, a natural candidate for such an investigation 
is the inspiral-merger-ringdown consistency test (IMRCT)~\cite{Ghosh:2016qgn, Ghosh:2017gfp,Shaikh:2024wyn,Madekar:2024zdj} (see also Refs.~\cite{Pompili:2025cdc,Chiaramello:2025bhi} for related work exploring deviations in the ringdown). 
In the remainder of this study, we perform injections of our BBS waveforms into Gaussian noise and apply the IMRCT to assess its capacity to uncover inconsistencies with the BBH hypothesis. Although the strongest signal degeneracies are observed for the compact BBS, for completeness we also include the analysis of the fluffy \texttt{A147} binaries in our discussion.

\section{Act II: Inspiral-merger-ringdown consistency test} \label{sec:imr}
\subsection{The test}

\begin{figure*}[p]
    \centering
    \includegraphics[width=0.29\linewidth] {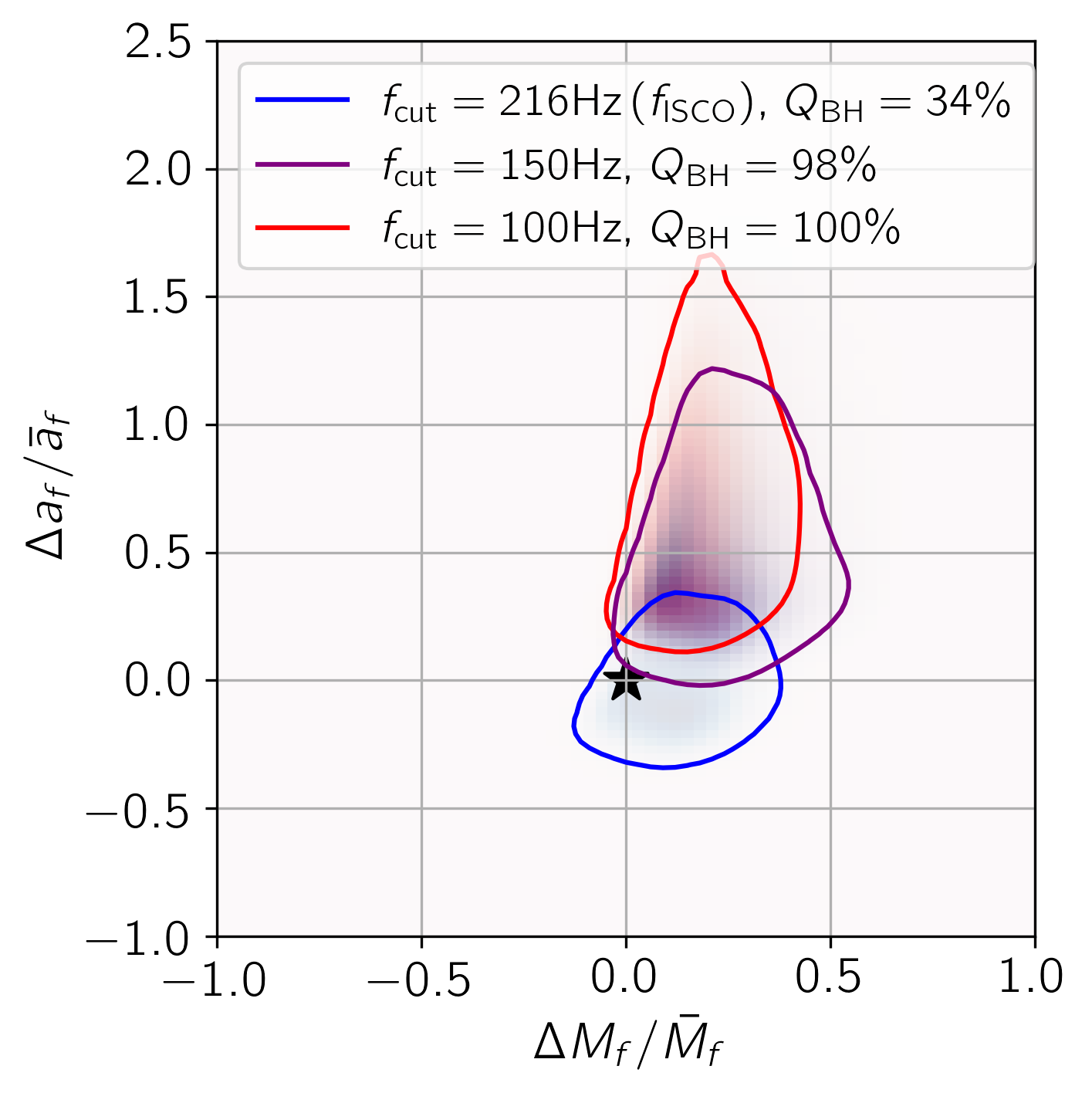} \quad
    \includegraphics[width=0.29\linewidth] {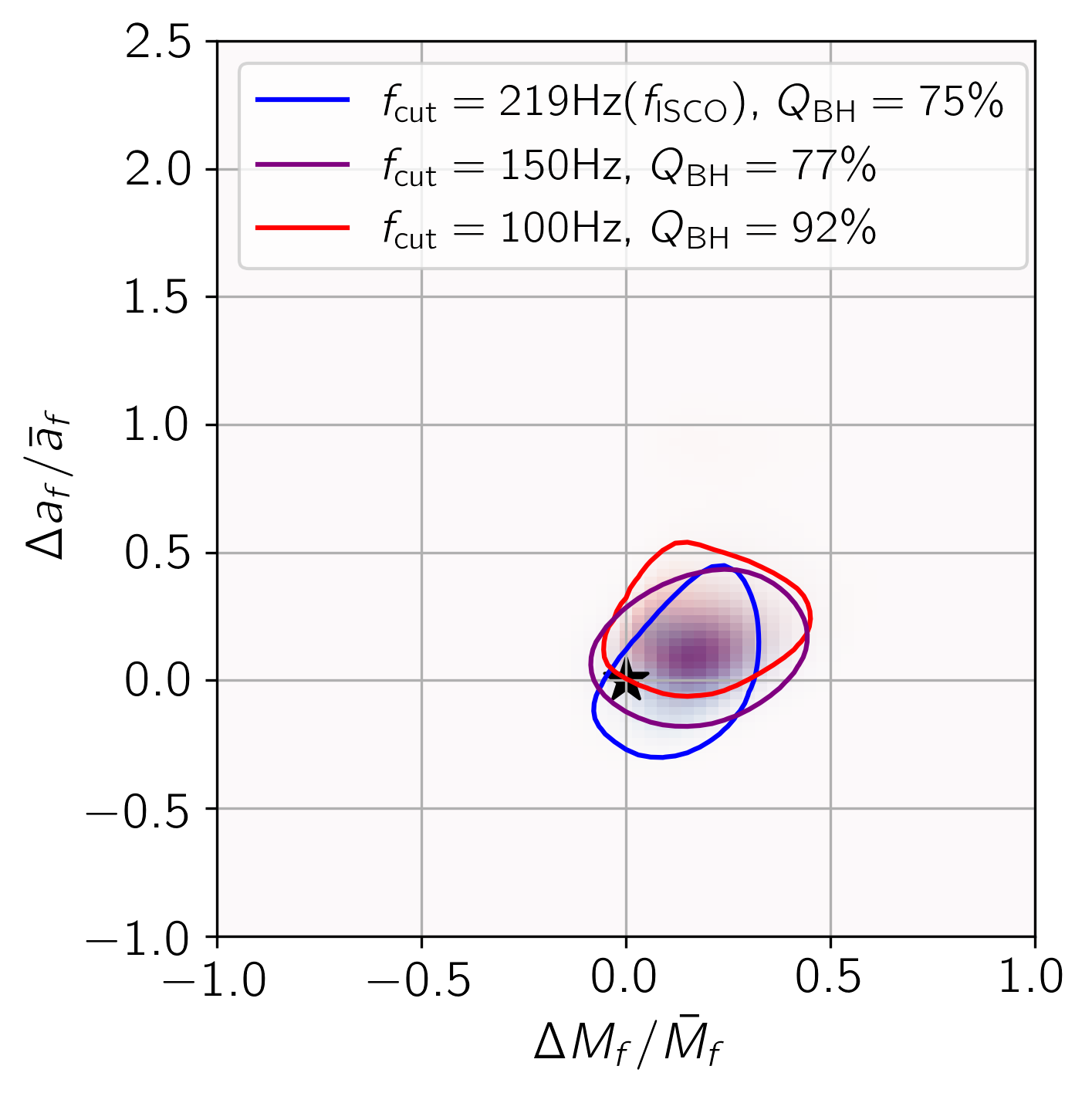}
    \\
    \includegraphics[width=0.29\linewidth] {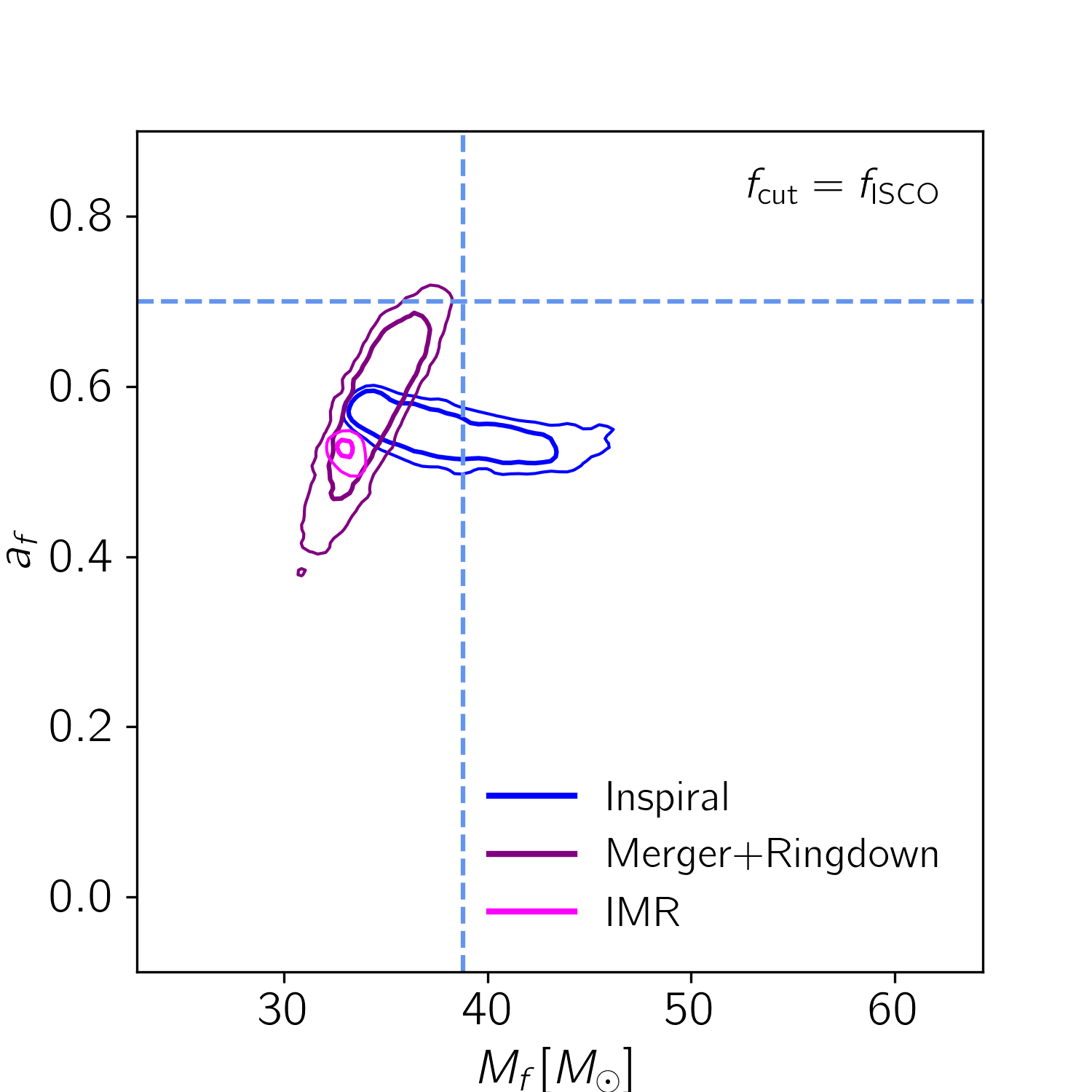} \quad 
    \includegraphics[width=0.29\linewidth] {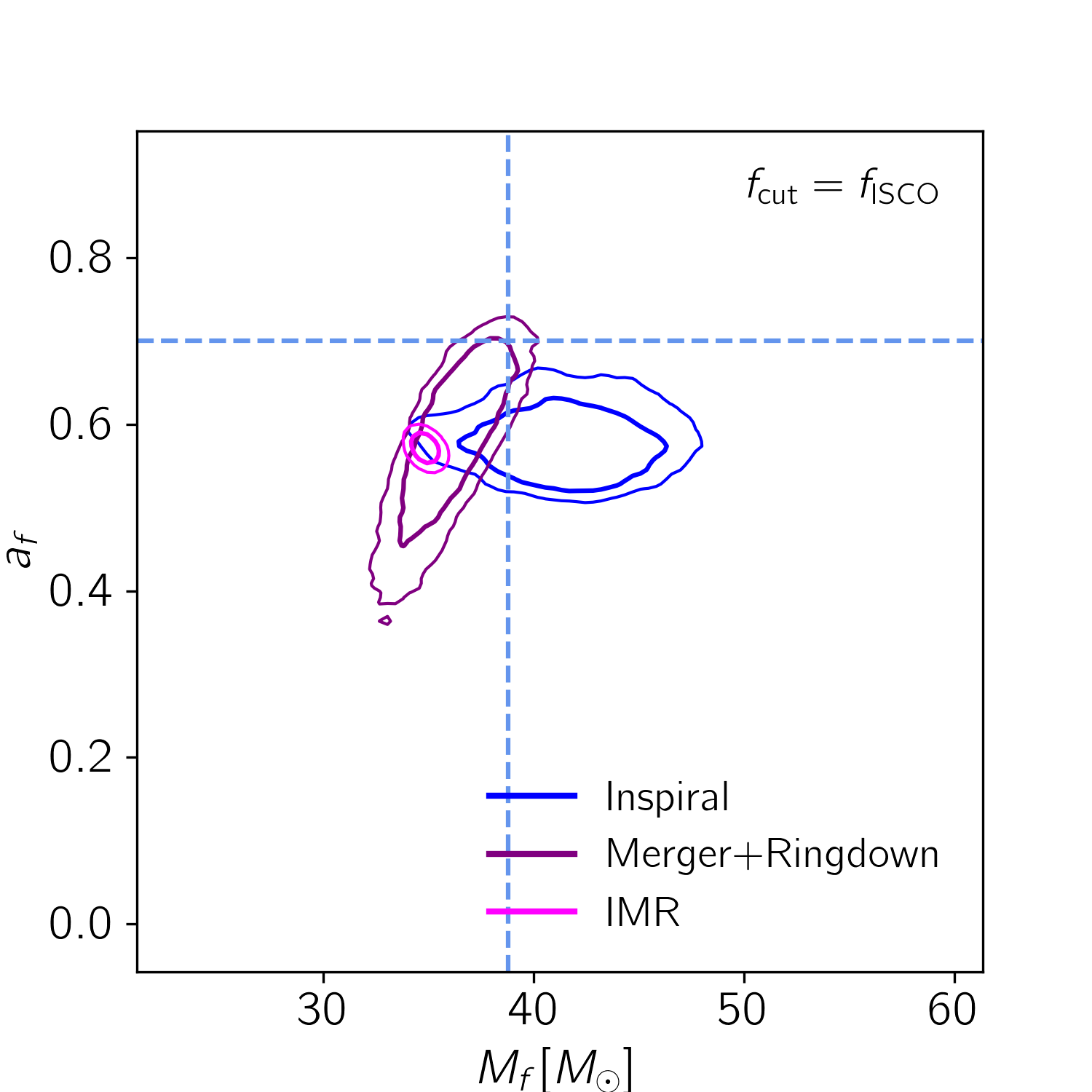}\\
    \includegraphics[width=0.29\linewidth] {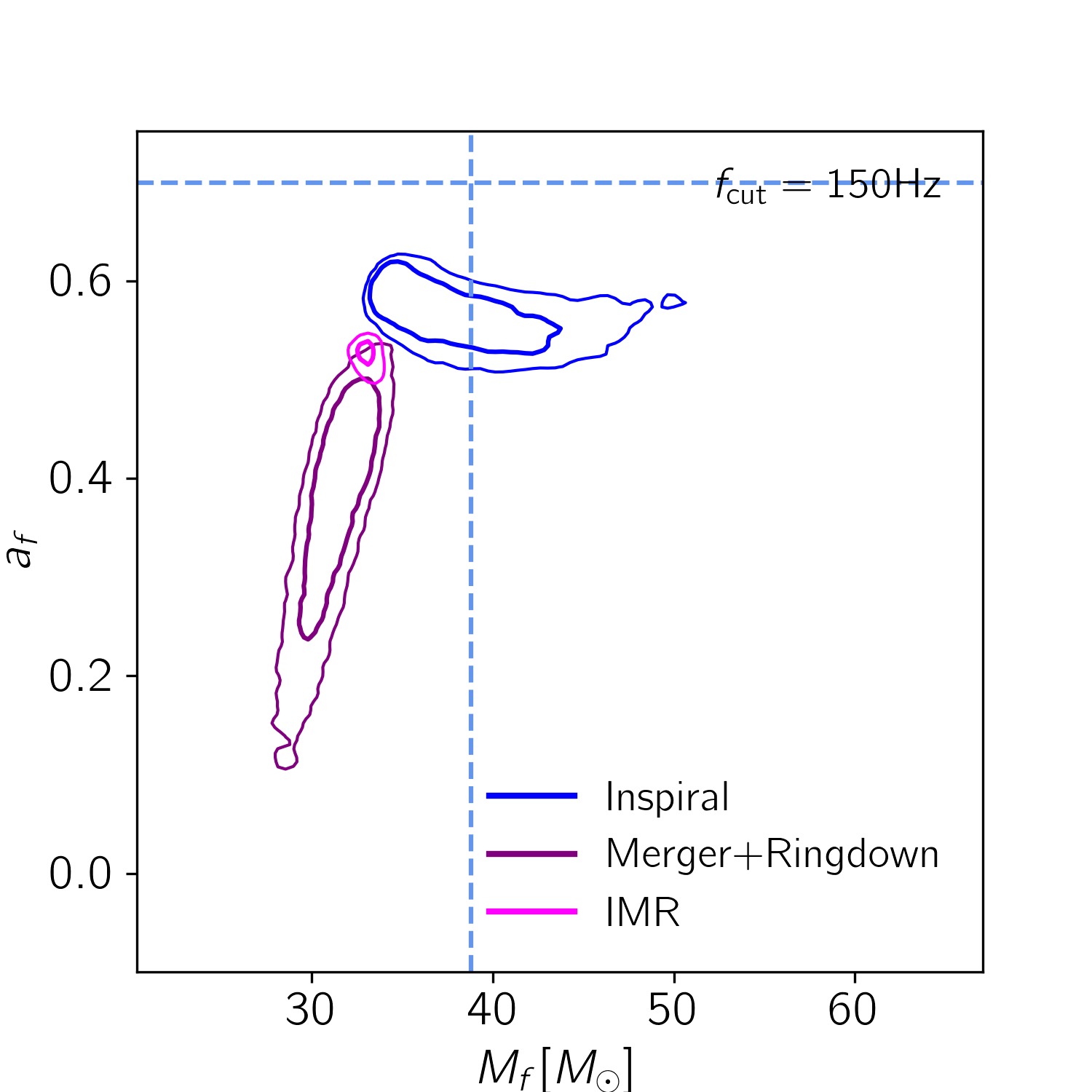} \quad
    \includegraphics[width=0.29\linewidth] {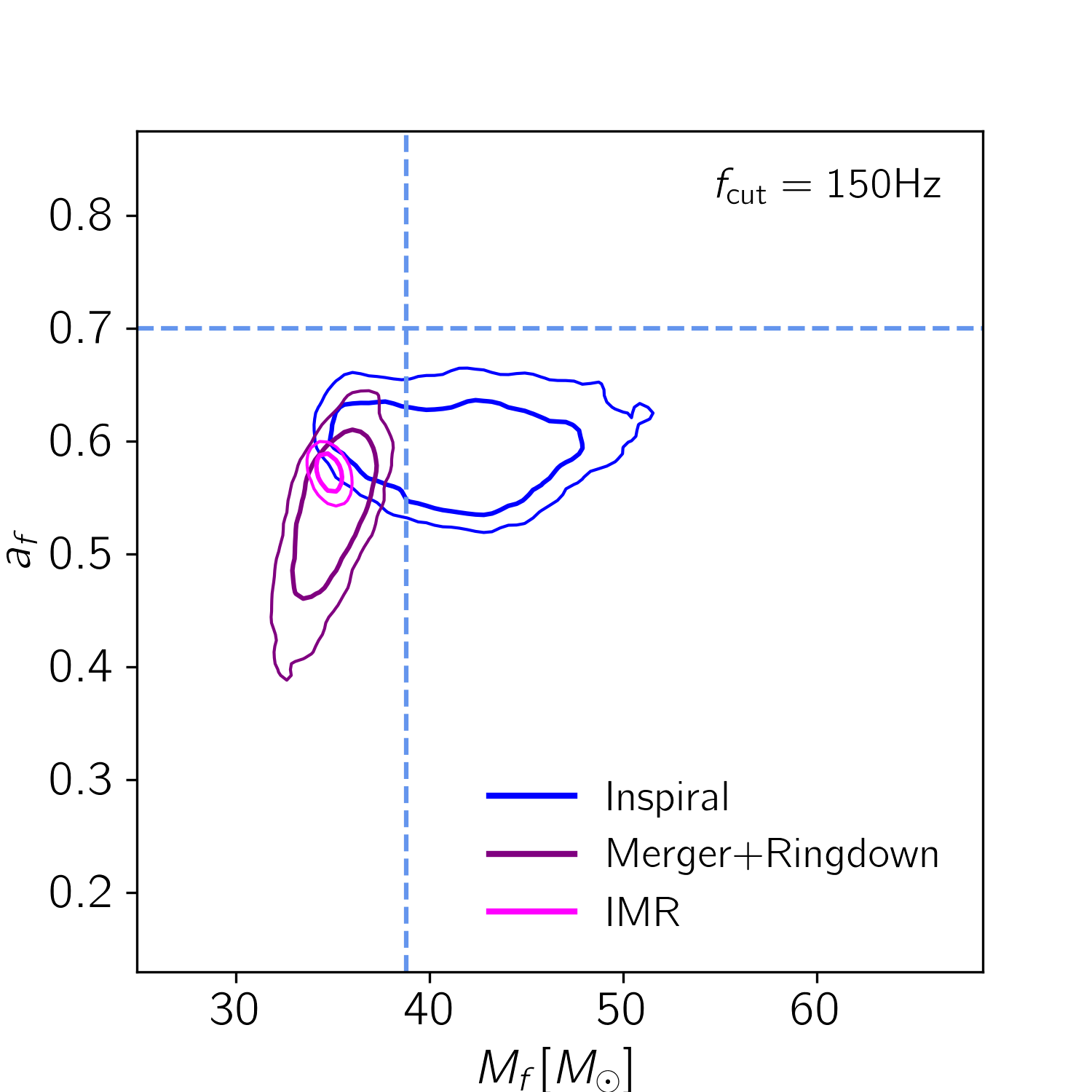} \\
    \includegraphics[width=0.29\linewidth] {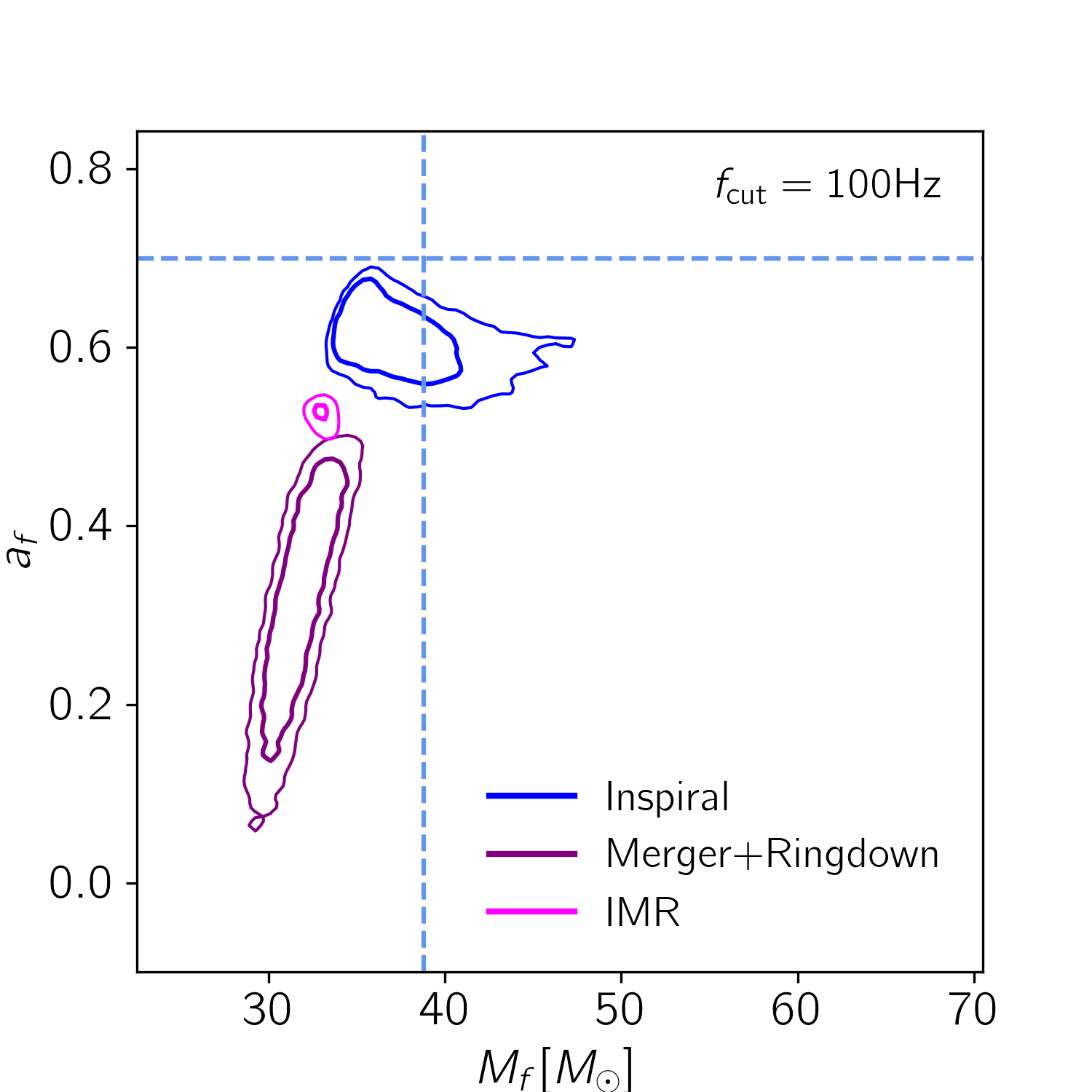} \quad
    \includegraphics[width=0.29\linewidth] {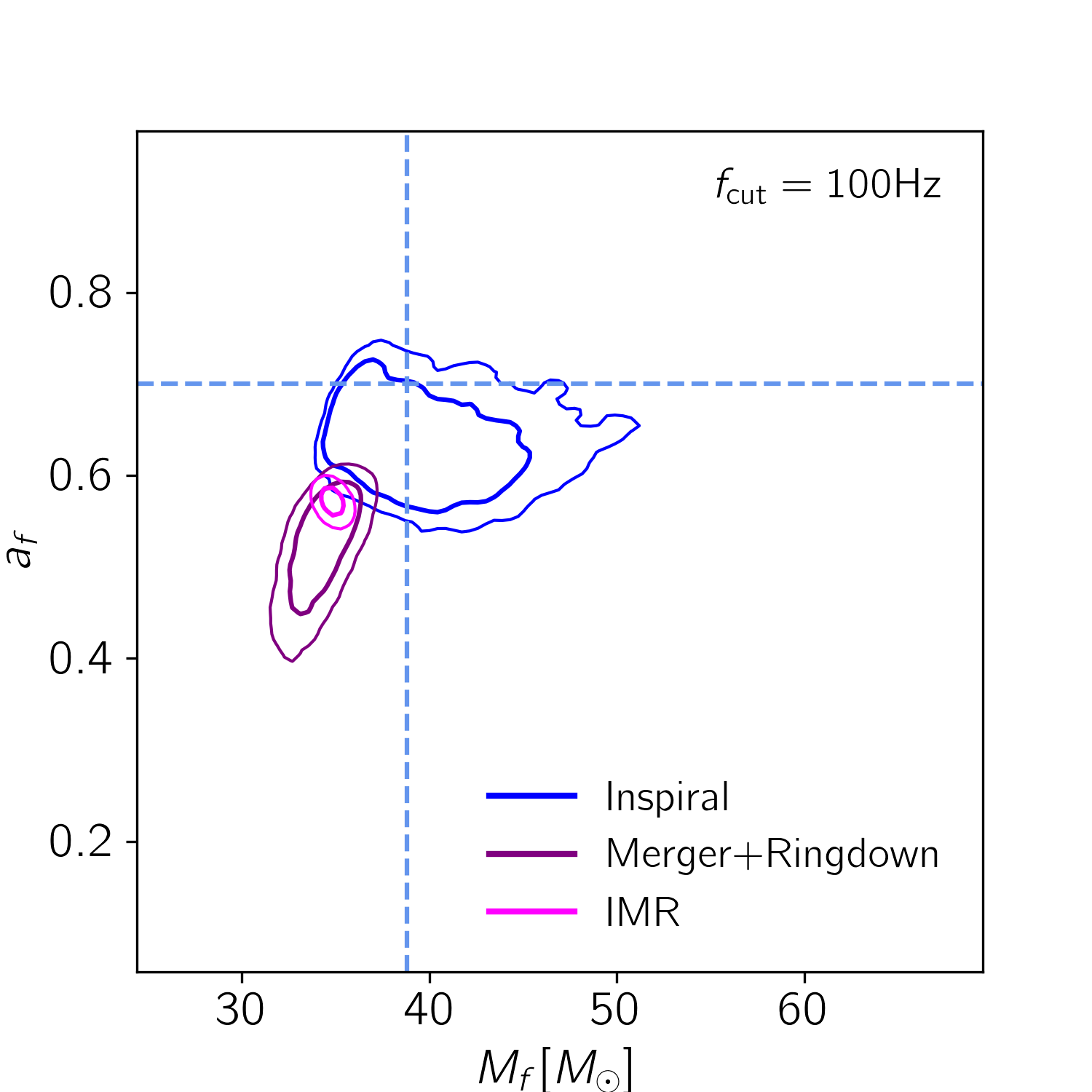}
    \caption{\textit{Left}: Results of the IMRCT analysis of the \texttt{A17-d15} binary with $M_{\rm{tot}} = 40M_{\odot}$, $d_L = 250 \rm{Mpc}$, $\rho_{\rm net} \sim 56$ and recovery with \texttt{IMRPhenomXAS}. In the top panel we plot the 90\% credible region of the $P(\epsilon, \sigma)$ posterior for various cut-off frequencies, $f_{\rm{cut}}$. In the lower three panels we show the behaviour of the 68\% and 90\% credible regions of $(M_{\rm f}, a_{\rm f})$ posteriors for the inspiral portion of the signal, the merger-ringdown part and, for reference, the posteriors obtained using the full IMR signal. The light blue dashed lines indicate the injected values. \textit{Right}: The same as the left column, but for the \texttt{IMRPhenomXP} waveform approximant with spin precession. 
    }
    \label{fig:imr-result}
\end{figure*}

The goal of the IMRCT, first proposed in Refs.~\cite{Ghosh:2016qgn, Ghosh:2017gfp}, is to perform parameter estimation on the low and high frequency chunks of the signal and check for consistency across the inferred parameters. Within the stationary phase approximation, the low (high) frequency portions of the signal correspond to the early (late) portions in the time domain. Therefore, the test can be loosely interpreted as splitting the signal into the inspiral (I) and merger-ringdown (MR) parts. However, the choice of where to split the signal in the frequency domain, $f_{\rm{cut}}$, is not unique. In particular, too high cut-off frequencies can introduce spectral leakage, where the ringdown can `leak' into the inspiral frequencies, while by using too small cut-off frequencies we risk losing too much information in the inspiral. Ref.~\cite{Ghosh:2017gfp} investigated various choices for the cut-off frequencies, up to the innermost-stable-circular-orbit frequency of the remnant Kerr BH, $f_{\rm ISCO}$, and found that frequencies $f \lesssim f_{\rm ISCO}$ robustly demarcate the signal into the `early' and `late' parts. The study also found that the choice of $f_{\rm{cut}}$ 
does not have much of an impact on the IMRCT 
for the type of deviations one may expect for BBHs in modified theories of gravity. Subsequently, many studies have since adopted the use of $f_{\rm ISCO}$ as the separating frequency between the inspiral and merger-ringdown~\cite{Ghosh:2016qgn, Ghosh:2017gfp,Shaikh:2024wyn,Madekar:2024zdj}. 

The consistency of the parameters in IMRCT is typically checked for by focusing on the posteriors of the final spin $a_{\rm f}$ and mass $M_{\rm f}$ of the remnant, inferred across the different parts of the signal (inspiral, merger-ringdown and full IMR). The null hypothesis assumes that these different parts of the signal belong to a BBH system, and $(a_{\rm f}, M_{\rm f})$ are therefore inferred from the progenitor masses and spins using analytical fits from BBH numerical relativity simulations~\cite{Hofmann:2016yih,Healy:2016lce,Jimenez-Forteza:2016oae} (specifically, we use fits provided in Ref.~\cite{Healy:2016lce}). 
Any inconsistency across these posteriors would imply that some parts of the signal are not described well by the waveform approximant in question. Typically, the distinction between the inspiral and merger-ringdown parts is quantified using \textit{fractional deviation parameters}
\begin{align}
    \epsilon \defeq \frac{\Delta M_{\rm f}}{\bar{M}_{\rm f}} &= 2 \frac{M_{\rm f}^{\rm I} - M_{\rm f}^{\rm MR}}{M_{\rm f}^{\rm I} + M_{\rm f}^{\rm MR}},
    \label{eq:epsilon} \\
    \sigma \defeq \frac{\Delta a_{\rm f}}{\bar{a}_{\rm f}} &= 2 \frac{a_{\rm f}^{\rm I} - a_{\rm f}^{\rm MR}}{a_{\rm f}^{\rm I} +a_{\rm f}^{\rm MR}}. 
\end{align}
The extent of the inconsistency is conveniently measured by computing the BH quantile, 
which quantifies the fraction of the posterior probability contained within the isocontour that intersects the BH 
value\footnote{We follow the same definition as the `GR quantile' used in the IMRCT literature, but adopt the BH quantile nomenclature in this paper, as we apply the IMRCT framework to BBS systems within GR, thereby testing their consistency with the BBH hypothesis.}  $(0,0)$

\begin{equation}
    Q_{\rm BH} \defeq \int_{P(\epsilon, \sigma) > P(0,0)} P(\epsilon, \sigma) \du \epsilon \du \sigma.
\end{equation}
In the remainder of the text, we will report the deviation to be significant whenever the BH quantile exceeds 90\%, however we will also quote its exact value in our discussions. 

\subsection{Set-up}

In our investigations, we focus on binary (detector-frame) masses in the range\footnote{We find the results of IMRCT discussed here qualitatively similar across the entire considered mass range.} $M_{\rm{tot}} \in [40, 80]M_{\odot}$, because (i) too low binary masses significantly increase the starting frequency of our NR injections
above $20 \rm{Hz}$ and considerably push the higher frequency content of our waveforms beyond the minimum of the detectors' sensitivity curves,
and (ii) too high binary masses significantly suppress the inspiral portion of the signal in the LVK band. The starting frequency of our analysis of course depends on the length of our simulations. For the range of $M_{\rm tot}$ we consider here, the largest starting frequency is $\sim 37 \rm Hz$ for the \texttt{A17-d12} family of simulations and $M_{\rm tot} = 40 M_{\odot}$.

In recovery we use the \texttt{IMRPhenomXP}~\cite{Pratten:2020ceb} waveform approximant, which includes the effect of spin precession and \texttt{IMRPhenomXAS}~\cite{Pratten:2020fqn}, which uses (anti-)aligned spins only. Our inference is performed in \textsc{bilby}~\cite{Ashton:2018jfp,nessai,Williams:2021qyt} and the employed set-up is the same as described in Ref.~\cite{Evstafyeva:2024qvp}, except that we now choose to inject into a network of three detectors, including Hanford, Livingston and Virgo. Our sky location parameters are chosen arbitrarily (but to avoid singular scenarios such as edge-on orientation) and fixed to right ascension $\alpha = 1.375$, declination $\delta = -1.2108$ and inclination angle $\iota = 60^{\degree}$. Most of the binaries considered in this work have moderate network SNRs $\rho_{\rm net} \sim 46-56$, well below the SNR threshold $\rho_{\rm max}$ determined by the NR waveform accuracy requirements discussed in Section~\ref{sec:mismatch}. In this total-SNR range, the $l=3$ ($l=4$) modes discussed in Section~\ref{sec:merger} have individual network SNRs of $\lesssim 1$ ($\lesssim 4$). While the power in the $l=4$ mode is more significant and could be detected in a realistic GW scenario~\cite{Mills:2020thr}, the accuracy of our NR simulations would need to be increased in the early inspiral.
We therefore focus on injecting the $l=2$ mode content of our waveforms.

\subsection{Results}

We start our discussion with the results of the IMRCT for compact \texttt{A17} binaries. We illustrate the key takeaways for the \texttt{A17-d15} configuration, and discuss the sensitivity of IMRCT on (i) $f_{\rm cut}$, (ii) the waveform approximant used in recovery (i.e.~precession vs aligned-spins only), and (iii) network SNR. The same picture holds for almost all other compact binary systems of Table~\ref{tab:waveforms},
with some exceptions for the anti-BS and $\delta \phi = \pi/2$ binaries which we will detail below. 

\subsubsection{A representative example: the in-phase \texttt{A17-d15} binary}

Our representative injection of the \texttt{A17-d15} binary with $M_{\rm{tot}} = 40 M_{\odot}$ and $d_{\rm{L}} = 250 {\rm Mpc}$ results in a network SNR $\rho_{\rm net} \approx 56$. In the left column of Figure~\ref{fig:imr-result}, we illustrate the dependency of IMRCT on $f_{\rm cut}$ using the aligned-spins waveform approximant \texttt{IMRPhenomXAS} in recovery\footnote{Unlike conventions of e.g.~Ref.~\cite{LIGOScientific:2026qni}, we do not reweigh the posteriors to impose uniform priors on the fractional deviation parameters $\epsilon$ and $\sigma$. For the majority of cases considered here, we have verified that reweighting the posterior samples does not qualitatively alter the IMRCT results.}. We note that the biased recovered parameters also bias our estimates of the final mass and spin obtained from the IMR signal; we obtain $M_{\rm f} \sim 33.3 M_{\odot}$ 
and $a_{\rm f} \sim 0.53$ as the median values in this case and $f_{\rm{ISCO}} \sim 216$\,Hz. This is to be contrasted with the final mass and spin estimates from the energy-balance arguments in Table~\ref{tab:models}: $M^{\rm inj}_{\rm f} \sim 38.8 M_{\odot}$ and $a^{\rm inj}_{\rm f} \sim 0.7$.
For $f_{\rm cut} = f_{\rm{ISCO}}$, IMRCT is insensitive to the BBS nature of this injection,
producing a BH quantile of $Q_{\rm BH}\sim 34\%$. From the left, second from top panel of Fig.~\ref{fig:imr-result}, one may 
already notice that the inspiral and merger-ringdown 
posteriors tend to have elongated shape in nearly orthogonal directions.
Despite this effect, however, the distributions retain sufficient overlap
and therefore do not
produce a parameter inconsistency.
In the time 
domain language of Fig.~\ref{fig:pn_match}, higher cut-off 
frequencies (including $f_{\rm cut} = f_{\rm{ISCO}}$) allow 
both the inspiral and merger-ringdown parts of the signal to 
contain regions where this particular equal-mass non-spinning 
BBS starts to differ from an equal-mass non-spinning BBH. IMRCT's inability to detect a parameter inconsistency in this setup is therefore not surprising.

We can also understand this behaviour by considering the PE with \texttt{IMRPhenomXAS} using the {\it full} IMR waveforms of \texttt{A17-d15}. That recovery indirectly recognizes the non-BBH character of the signal by inferring
asymmetric mass ratios and non-zero anti-aligned spins, at 
least for the primary. When splitting up the signal at
sufficiently high $f_{\rm cut}$ this non-BBH nature of the signal
is more or less evenly split between the inspiral and the merger-ringdown parts, leaving the two on `average' consistent with one 
another.
By lowering the cut-off frequency, in contrast, we slowly approach the regime where the inspiral part of our BBS waveform agrees with the PN predictions for BBHs. Thus, restricting the inspiral analysis to ever earlier parts where BH and BS binaries generate nearly identical wave signals, the BBH approximant naturally recovers the correct injected masses and spins. The merger-ringdown part, however, now contains almost all peculiarities of the BS signal and yields correspondingly biased parameter estimates. In consequence, we see significant inconsistency in the inferred final masses and spins in the inspiral and merger-ringdown parts. This expectation is corroborated by the bottom two panels in the left column of Fig.~\ref{fig:imr-result}: IMRCT clearly picks up discrepancies in the $(M_{\rm f}, a_{\rm f})$ parameters for cut-off frequencies $f_{\rm cut} \leq 150 \rm{Hz}$, where the BH quantile of the fractional deviation parameters becomes larger than 90\%, showing clear evidence against the null hypothesis.

By using the aligned-spin approximant \texttt{IMRPhenomXAS}
in the above analysis, we are excluding the degrees of freedom
in the GW signals associated with spin precession.
This limitation
reduces the approximant's capability of reproducing the
``renegade'' BS waveform. 
The inclusion of precession, on the other hand, introduces more degrees of freedom and \textit{enhances} the degeneracy between BBS and BBH waveforms. Therefore, to test the efficacy of IMRCT in a more challenging scenario, we repeat the the above analysis, but now using the \texttt{IMRPhenomXP} waveform approximant with precession effects turned on. We report our results in the right column of Fig.~\ref{fig:imr-result}. In line with our previous discussion, we find that splitting the waveform at $f_{\rm cut} = f_{\rm ISCO}$ disguises the inconsistencies in the inferred parameters for the IMRCT. Decreasing $f_{\rm cut}$ still enables us to uncover the inconsistency in the $P(\epsilon, \sigma)$ posteriors. However, the effects of precession tend to smear out the drastic differences between the inspiral and merger-parts of the signal and typically reduce the BH quantiles compared to the aligned-spins recoveries. A confident identification of nonzero fractional deviation parameters is obtained only by lowering $f_{\rm cut}$ to $100\,{\rm Hz}$ where the BH quantile reaches $92\,\%$. 

Finally, we comment on the sensitivity of IMRCT to the network SNR. As we have seen above for our fiducial network SNR $\rho_{\rm net} \sim 56$, IMRCT is unable to find an inconsistency in inspiral and merger-ringdown parts for $f_{\rm cut} = f_{\rm{ISCO}}$. Keeping $f_{\rm cut} = f_{\rm{ISCO}}$ and increasing the network SNRs to $\rho_{\rm net} \gtrsim 80$, however, we find that IMRCT starts to produce BH quantiles $\gtrsim 90 \%$; see Fig.~\ref{fig:snr-grq}. Therefore, IMRCT may already provide robust results at $f_{\rm cut} = f_{\rm{ISCO}}$ for `golden' binaries, at least, with deviations akin to in-phase binaries like \texttt{A17-d15}. 

\begin{figure}[t!]
    \includegraphics[width=0.95\linewidth]{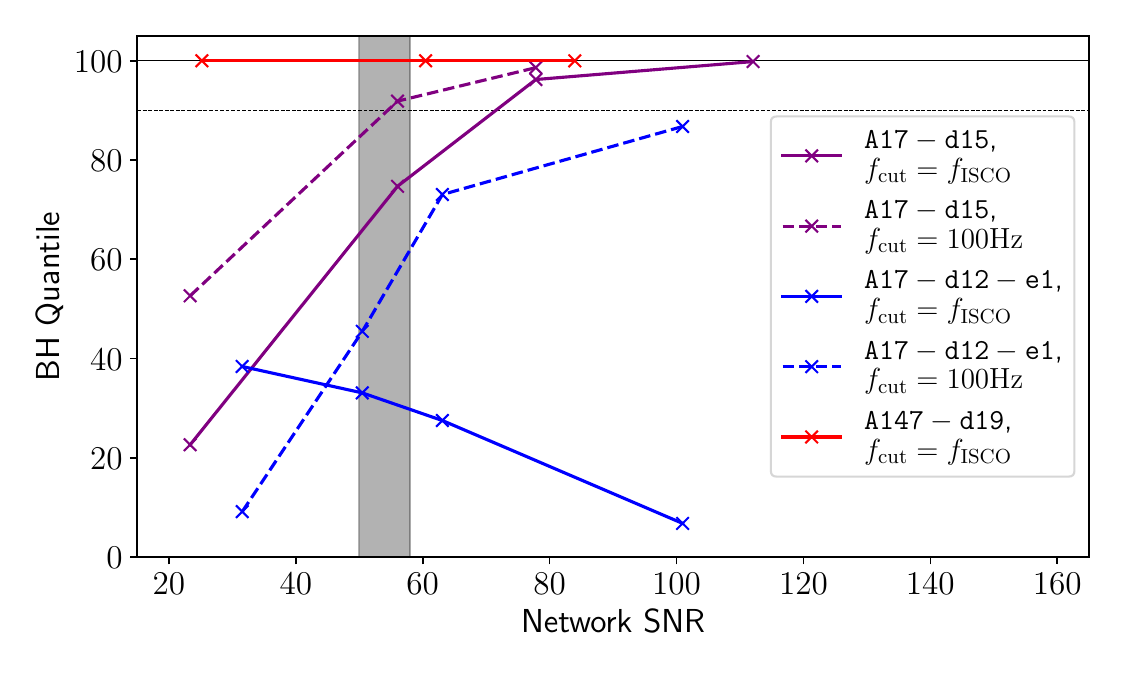} 
    \caption{
    The BH quantile as a function of network SNR for the \texttt{A17-d15}, \texttt{A17-d12-e1} and \texttt{A147-d19} binaries (with $M_{\rm tot} = 40 M_{\odot}$) at two representative cut-off frequencies, $f_{\rm cut} = f_{\rm ISCO}$ and $f_{\rm cut} = 100 \rm{Hz}$. The results presented here utilise \texttt{IMRPhenomXP} for recovery. The horizontal lines mark BH quantiles of $90\%$
    and $100\%$. Note that for compact binaries lower cut-off frequencies generally result in larger BH quantiles. While for the \texttt{A17-d15} binary we find inconsistency at lower $f_{\rm cut}$ and moderate SNR or at higher SNR and any $f_{\rm cut}$, the anti-BS binary \texttt{A17-d17-e1} passes the IMRCT throughout. Contrary to that, IMRCT confidently uncovers an inconsistency in the fluffy binary \texttt{A147-d14} even for $f_{\rm cut} = f_{\rm ISCO}$. The shaded region marks the range of moderate network SNR considered in Figs.~\ref{fig:imr-result} and \ref{fig:deviation-p120-p180}.
    }
    \label{fig:snr-grq}
\end{figure}

\begin{figure}[t!]
    \includegraphics[width=0.9\linewidth]{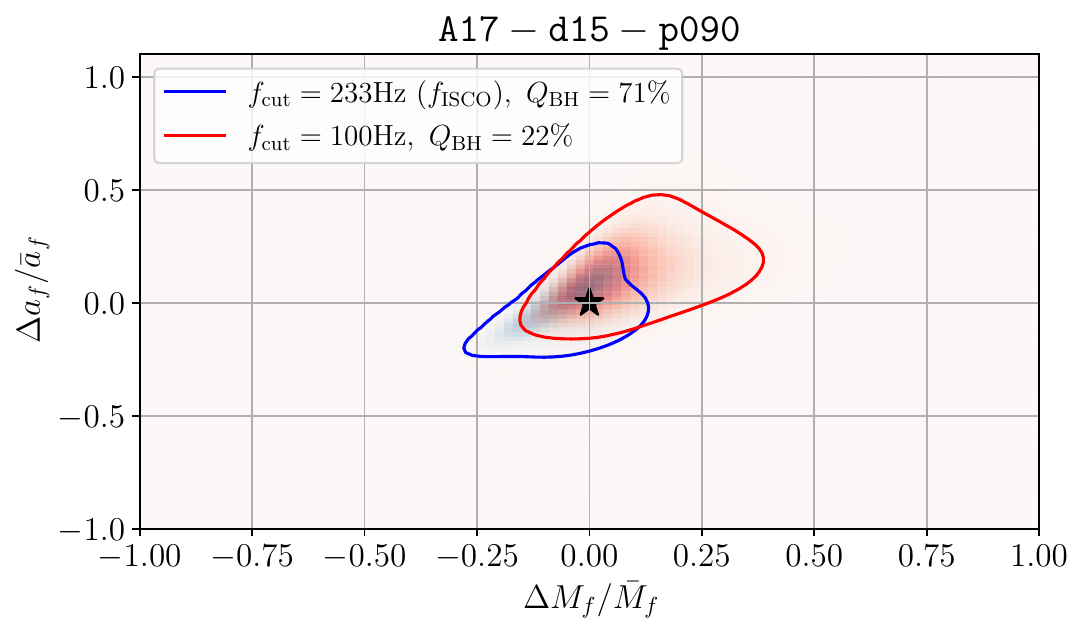}
    \includegraphics[width=0.9\linewidth]{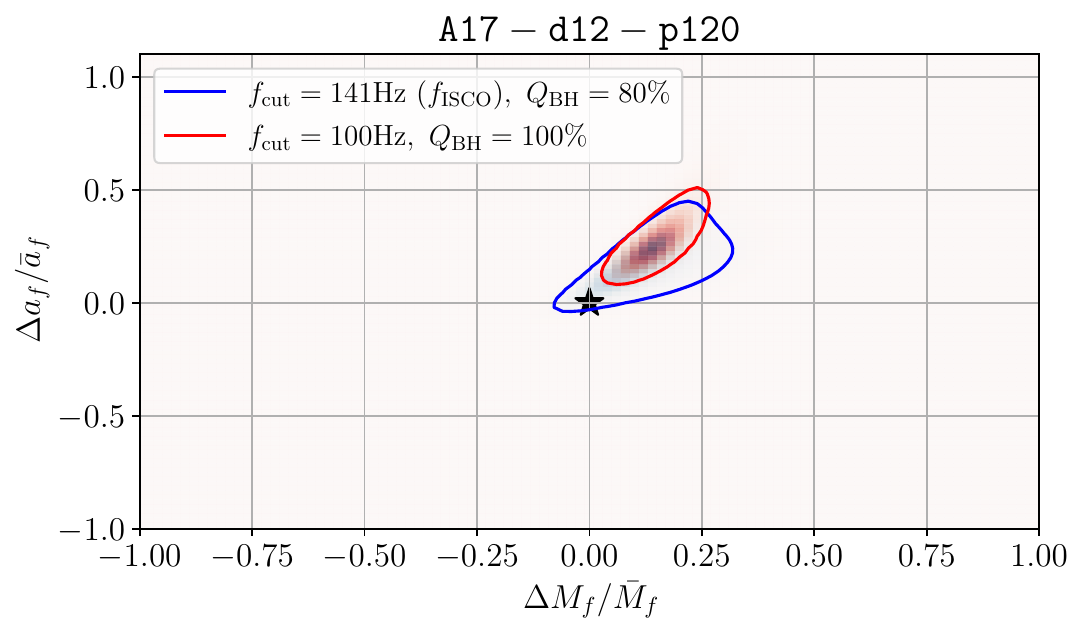}
    \includegraphics[width=0.9\linewidth]{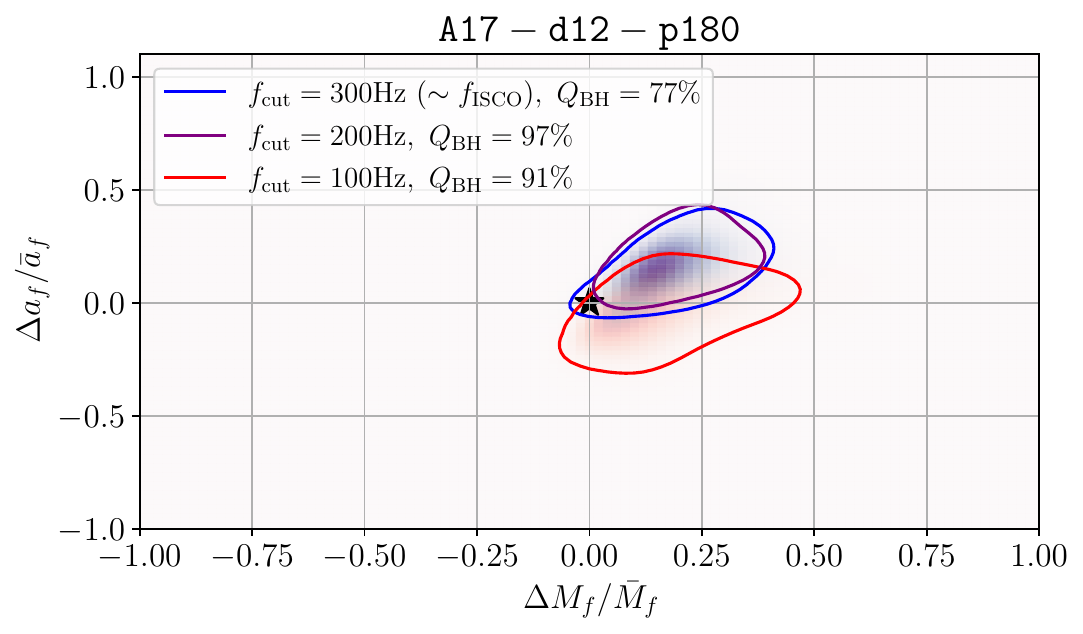}
    \caption{Results of the IMRCT for the \texttt{A17-d15-p090} injection with $M_{\rm tot} = 40 M_\odot$, $d_L = 250  {\rm Mpc}$ and $\rho_{\rm net} \sim 58$ (top); the \texttt{A17-d12-p120} injection with $M_{\rm tot} = 80 M_\odot$, $d_L = 500  {\rm Mpc}$ and $\rho_{\rm net} \sim 50$ (center); and the \texttt{A17-d12-p180} injection with $M_{\rm tot} = 40 M_\odot$, $d_L = 250  {\rm Mpc}$ and $\rho_{\rm net} \sim 53$ (bottom). We plot the 90\% credible region of the $P(\epsilon, \sigma)$ posterior for various cut-off frequencies, $f_{\rm{cut}}$, utilising the \texttt{IMRPhenomXP} waveform approximant in recovery.}
    \label{fig:deviation-p120-p180}
\end{figure}

\subsubsection{Compact dephased binaries}

Having discussed the results of IMRCT on the representative in-phase binary \texttt{A17-d15}, let us now focus on the other compact configurations: dephased, anti-phase and anti-BS binaries. 

Firstly, the anti-phase ($\delta \phi = \pi$) and $\delta \phi = 2\pi/3$ binaries produce qualitatively similar results to the representative example we have presented in the previous section. In the bottom two panels of Fig.~\ref{fig:deviation-p120-p180} we illustrate the behavior of $P(\epsilon, \sigma)$ with $f_{\rm cut}$ at moderate network SNR. Similar to the in-phase binary, the BH quantile increases with lower $f_{\rm cut}$. This behaviour is expected, since the most drastic deviations in $\delta \phi = \pi$ and $\delta \phi = 2\pi/3$ binaries from an equal-mass non-spinning BBH originate in the late inspiral and merger.

In contrast, for the anti-BS binary, we find the IMRCT to vary quantitatively with variations in $f_{\rm cut}$ but not to confidently spot an inconsistency for a wide range
of network SNRs (cf.~Fig.~\ref{fig:snr-grq}). Only for very large network SNR ($\rho_{\rm net} \geq 100$), we find that the BH quantile has a tendency to increase more significantly towards 90\% at lower $f_{\rm cut} = 100 \rm{Hz}$. This relative `insensitivity' of IMRCT likely arises because the anti-BS binary more closely resembles an equal-mass non-spinning BBH in both, the inspiral and merger parts of the signal (due to averaging out of short-range scalar interactions~\cite{Palenzuela:2006wp,Evstafyeva:2024qvp}). Similarly, as illustrated in the top panel of Fig.~\ref{fig:deviation-p120-p180}, IMRCT is insensitive to $f_{\rm cut}$ for binaries with $\delta \phi = \pi/2$ at moderate network SNR of $\rho _{\rm net} \sim 58$. However, analyses of these binaries tend to produce larger BH quantiles compared to anti-BS configurations. Furthermore, at large network SNRs $\rho_{\rm net} \gtrsim 80$, the parameter inconsistency in the inspiral and merger-ringdown parts of $\delta \phi = \pi/2$ binary becomes significant enough to produce BH quantiles exceeding $90 \%$ at $f_{\rm cut} = f_{\rm ISCO}$.

\subsubsection{Fluffy binary}

Finally, we apply the IMRCT to the fluffy \texttt{A147} binaries. As expected, for these types of binaries the test is efficient in revealing the parameter inconsistencies between inspiral and merger-ringdown parts even with the most general \texttt{IMRPhenomXP} approximant. Already at $f_{\rm cut} = f_{\rm ISCO}$ we find BH quantiles of 100\%.
As can be seen in Fig.~\ref{fig:snr-grq}, even at lower SNRs $\rho_{\rm net} \sim 25$, IMRCT successfully picks up these inconsistencies. 

\section{Conclusions}

In this work we have analysed the GW phenomenology of equal-mass non-spinning inspiralling boson star binaries using numerical relativity simulations. Our study focused on two representative systems whose progenitors differ in compactness: (i) fluffy configurations with compactness $C\approx 0.1$  and (ii) compact configurations with $C\approx 0.2$. We summarise the key findings regarding their GW morphology as follows:
\begin{list}{\rm{(\arabic{count})}}{\usecounter{count}
             \labelwidth0.5cm \leftmargin0.7cm \labelsep0.2cm \rightmargin0cm
             \parsep0.5ex plus0.2ex minus0.1ex \itemsep0ex plus0.2ex}
    \item The early inspiral of both systems is well described by 3.5PN estimates for equal-mass non-spinning BBHs. 
    \item The late inspiral and merger are strongly affected by the initial dephasing $\delta \phi$ and the sign $\epsilon$ of the scalar frequency. These parameters not only impact the shape of the chirp of the signal~\cite{Evstafyeva:2024qvp}, but also lead to the excitation of odd $m$ harmonics in binaries with $\delta \phi ~\mathrm{mod}~ \pi \ne 0$, although these modes are a few orders of magnitude smaller than the dominant $(lm)=(22)$ contribution. Nonetheless, these observations underpin the importance of short-range interactions of the scalar fields around merger, as is also observed at close encounter in scattering
    BS binaries \cite{Damour:2025oys}.
    \item For compact binaries whose remnants collapse to BHs post-merger, the late ringdown frequencies are similar to those of Kerr BHs, estimated from the final masses and spins of the remnants. 
    The largest deviations occur in the damping times, reaching at most  $\sim 13 \%$ for in-phase ($\delta \phi = 0$) binaries. On the contrary, fluffy configurations forming a BS remnant, exhibit much longer damping times and smaller real frequencies. 
    \item For both, compact and fluffy BS binaries, we can model the late ringdown reasonably well with a single damped sinusoid. We also find tentative evidence for a second mode, possibly an overtone, although its extracted frequencies show less stability with respect to the choice of ringdown start time when compared to the fundamental mode. 
\end{list}
From the above summary, it is clear that late inspiral and merger parts of compact BBS systems offer the most promising avenues towards differentiating them from BBHs.

In light of this we have performed inspiral-merger-ringdown consistency tests to determine whether deviations from the BBH hypothesis can be identified. Overall, at moderate network SNR ($\rho_{\rm net} \lesssim 70$) we find the IMRCT with the fiducial $f_{\rm cut} = f_{\rm ISCO}$, the separating frequency between the inspiral and merger–ringdown parts of the signal, to be insufficient for uncovering parameter inconsistencies between compact BS and BH binaries. Depending on the concrete example of such compact BS binaries, parameter inconsistencies can typically be found at the fiducial $f_{\rm cut}$ for very large SNR events or by reducing $f_{\rm cut}$ and thus restricting the inspiral part of the signal.
In particular, deviations from the BBH hypothesis were observed at 90\% confidence level only for $f_{\rm cut} < f_{\rm ISCO}$, with some exceptions for the anti-BS and $\delta \phi = \pi/2$ binaries (cf.~Table~\ref{tab:summary-imrct}).

For very loud injections with $\rho_{\rm net} \gtrsim 80$, however, the separating frequency of $f_{\rm cut} = f_{\rm ISCO}$ was sufficient to reveal the non-BBH origin of all BBS waveforms, except for the anti-BS binary. Unlike other BBS configurations, anti-BS binaries tend to produce GW signals that very closely resemble those of comparable BBHs~\cite{Evstafyeva:2024qvp}, thereby making IMRCT rather insensitive to them for various $f_{\rm cut}$ and a wide range of network SNRs ($\rho_{\rm net} \sim 20 - 100$).
These binaries therefore offer interesting examples where even moderately compact stars ($C \sim 0.2)$ can mimick GW signatures of a BBH system. 
Finally, for fluffy configurations, IMRCT with the fiducial choice of $f_{\rm cut} = f_{\rm ISCO}$ successfully identifies deviations from the BBH hypothesis for network SNRs as low as $\rho_{\rm net} \geq 25$; this is unsurprising bearing in mind that these systems were
already flagged by the comparatively simple residual test
\cite{Evstafyeva:2024qvp}.

\begin{table}[t!]
    \centering
    \begin{tabular}{l@{\hspace{0.2cm}}| @{\hspace{0.2cm}} c @{\hspace{0.5cm}} c}
        \toprule
        $f_{\rm cut}$ & $f_{\rm ISCO}$ & 100 Hz \\
        \midrule
        \texttt{A17-d15} & no & yes \\
        \texttt{A17-d15-p090} & no & no \\
        \texttt{A17-d12-p120} & no & yes \\
        \texttt{A17-d12-p180} & no & yes \\
        \texttt{A17-d12-e1} & no & no \\
        \texttt{A147-d19} & yes & yes \\
        \bottomrule
    \end{tabular}
    \caption{Summary of the sensitivity of IMRCT to various binary boson stars injections at network SNRs $\rho_{\rm net} \lesssim 70$ and cutoff frequencies $f_{\rm cut}$. A yes/no in the table refers to whether or not the IMRCT results in a BH quantile of at least $90\%$, quantifying the inconsistency in the inferred parameters.} \label{tab:summary-imrct}
\end{table}

We conclude by commenting on possible implications of the above discussions for future GW observations. Although in certain regions of the parameter space BBSs may `trick' IMRCT, a realistic astrophysical population of BSs spanning a range of masses and compactness would likely leave detectable imprints in the data. In particular, if a sub-population of BSs exists, it is plausible that it would also 
manifest itself at specific binary mass scales (set by the scalar mass $\mu$), where deviations from the BBH hypothesis are expected to be most pronounced. For example, such deviations 
can arise from less compact binaries ( akin to our \texttt{A147} family of runs), where IMRCT can yield significant BH quantiles, signaling a tension with the BBH paradigm. Therefore, the absence of such signatures in current GW observations can already place constraints on the properties of possible BS populations and potentially rule out the extreme scenario of BBSs replacing all BBH systems observed to the present date. 

\begin{acknowledgments}
T.E. thanks William East, Luis Lehner and Aditya Vijaykumar for useful discussions. Research at Perimeter Institute is supported in part by the Government of Canada through the Department of Innovation, Science and Economic Development and by the Province of Ontario through the Ministry of Colleges and Universities.
We acknowledge support by the NSF Grant Nos.~PHY-090003,~PHY-2513337
and AST-2307146, by the Cambridge Service for Data Driven Discovery
at the University of Cambridge and
Durham University through DiRAC Projects ACTP284 and ACTP238 and STFC capital Grants No.~ST/P002307/1 and
No.~ST/R002452/1, and STFC operations Grant No.~ST/R00689X/1. Authors are grateful to computational resources provided by the LIGO Laboratory and supported
by National Science Foundation Grants PHY0757058 and
PHY0823459 and the support provided by SciNet HPC Consortium. SciNet is funded by Innovation, Science and Economic Development Canada; the Digital Research Alliance of Canada; the Ontario Research Fund: Research Excellence; and the University of Toronto. Computations were done on the CSD3 (Cambridge), Cosma7 and 8 (Durham),  Niagara and Trillium (University of Toronto),
Stampede3 (TACC) and Expanse (SDSC),
and CIT LIGO Lab (Pasadena) clusters. 
\end{acknowledgments}

\appendix

\section{Two-mode ringdown fits}\label{app:two_mode_fits}

\begin{figure}[t!]
    \centering
    \includegraphics[width=0.8\linewidth]{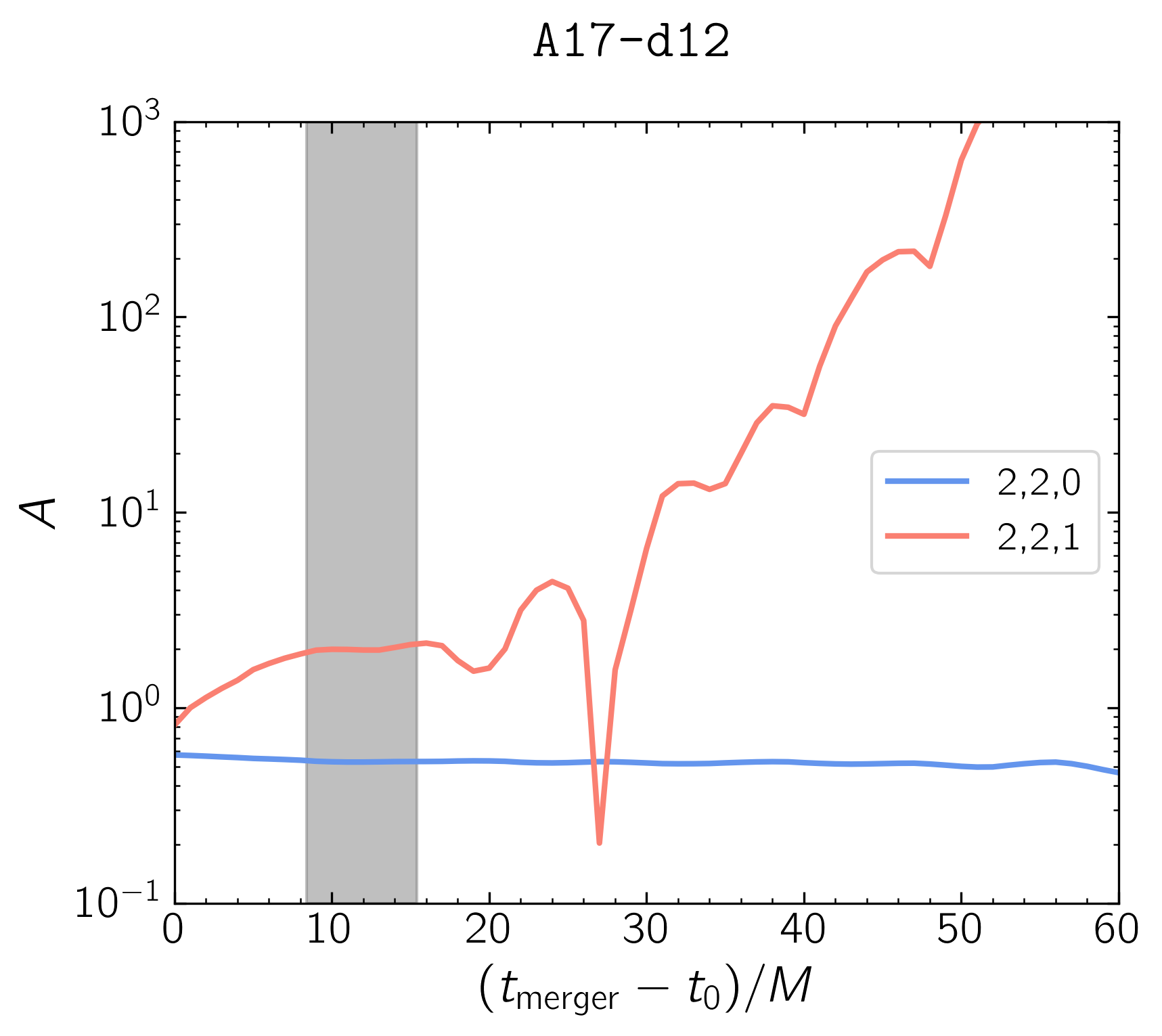}
    \caption{The amplitudes $A_{220}$ and $A_{221}$ of the fundamental and first overtone modes, as defined in Eq.~(\ref{eq:ringdown}), 
    are shown for the \texttt{A17-d12} binary. Here, we fix the complex frequencies at the values
    determined for a ringdown starting time $t_0=10\,M$, but compute
    the mode amplitudes for varying $t_0$.
    The shaded grey region shows the time window within which the amplitude of the second mode fluctuates by at most 10\% relative to its value for $t_{\rm merger} - t_0=10\,M$.
    }
    \label{fig:A17-A147-stability}
\end{figure}

As discussed in Section~\ref{sec:ringdown} of the main text, we can confidently identify only one mode in the ringdown parts of our BBS waveforms. More specifically, when fitting for two modes, $n(K)=2$, the second mode exhibits unsatisfactory stability: its frequency varies drastically with the chosen ringdown starting time $t_0$. 
For instance, for the \texttt{A17-d12} binary, its real frequencies (imaginary frequencies) change by $\sim 16\%$ ($\sim 11 \%$) in the window of ringdown starting times $t_0 \in [10, 15]M$.
Likewise, if we fix the mode frequencies to the values extracted for $t_0 = 10M$ and then monitor the mode amplitudes over time $(t_{\rm merger}-t_0)$, the amplitude of the second mode remains more or less stable for a relatively short while, but varies by several orders of magnitude at later times. We illustrate these variations in Figure~\ref{fig:A17-A147-stability} for the \texttt{A17-d12} binary. The behaviour of \texttt{A147} binaries is qualitatively similar.

\newpage

\bibliography{bibliography}

@article{Purrer:2019jcp,
    author = {P\"urrer, Michael and Haster, Carl-Johan},
    title = "{Gravitational waveform accuracy requirements for future ground-based detectors}",
    eprint = "1912.10055",
    archivePrefix = "arXiv",
    primaryClass = "gr-qc",
    doi = "10.1103/PhysRevResearch.2.023151",
    journal = "Phys. Rev. Res.",
    volume = "2",
    number = "2",
    pages = "023151",
    year = "2020"
}

@article{Evstafyeva:2024qvp,
    author = "Evstafyeva, Tamara and Sperhake, Ulrich and Romero-Shaw, Isobel M. and Agathos, Michalis",
    title = "{Gravitational-Wave Data Analysis with High-Precision Numerical Relativity Simulations of Boson Star Mergers}",
    eprint = "2406.02715",
    archivePrefix = "arXiv",
    primaryClass = "gr-qc",
    doi = "10.1103/PhysRevLett.133.131401",
    journal = "Phys. Rev. Lett.",
    volume = "133",
    number = "13",
    pages = "131401",
    year = "2024"
}

@article{London:2014cma,
    author = "London, Lionel and Shoemaker, Deirdre and Healy, James",
    title = "{Modeling ringdown: Beyond the fundamental quasinormal modes}",
    eprint = "1404.3197",
    archivePrefix = "arXiv",
    primaryClass = "gr-qc",
    doi = "10.1103/PhysRevD.90.124032",
    journal = "Phys. Rev. D",
    volume = "90",
    number = "12",
    pages = "124032",
    year = "2014",
    note = "[Erratum: Phys.Rev.D 94, 069902 (2016)]"
}

@article{Blanchet:2013haa,
    author = "Blanchet, Luc",
    title = "{Post-Newtonian Theory for Gravitational Waves}",
    eprint = "1310.1528",
    archivePrefix = "arXiv",
    primaryClass = "gr-qc",
    doi = "10.12942/lrr-2014-2",
    journal = "Living Rev. Rel.",
    volume = "17",
    pages = "2",
    year = "2014"
}

@article{Pan:2007nw,
    author = "Pan, Yi and Buonanno, Alessandra and Baker, John G. and Centrella, Joan and Kelly, Bernard J. and McWilliams, Sean T. and Pretorius, Frans and van Meter, James R.",
    title = "{A Data-analysis driven comparison of analytic and numerical coalescing binary waveforms: Nonspinning case}",
    eprint = "0704.1964",
    archivePrefix = "arXiv",
    primaryClass = "gr-qc",
    doi = "10.1103/PhysRevD.77.024014",
    journal = "Phys. Rev. D",
    volume = "77",
    pages = "024014",
    year = "2008"
}

@article{Siemonsen:2023hko,
    author = "Siemonsen, Nils and East, William E.",
    title = "{Binary boson stars: Merger dynamics and formation of rotating remnant stars}",
    eprint = "2302.06627",
    archivePrefix = "arXiv",
    primaryClass = "gr-qc",
    doi = "10.1103/PhysRevD.107.124018",
    journal = "Phys. Rev. D",
    volume = "107",
    number = "12",
    pages = "124018",
    year = "2023"
}

@article{Macedo:2013jja,
    author = "Macedo, Caio F. B. and Pani, Paolo and Cardoso, Vitor and Crispino, Lu\'\i{}s C. B.",
    title = "{Astrophysical signatures of boson stars: quasinormal modes and inspiral resonances}",
    eprint = "1307.4812",
    archivePrefix = "arXiv",
    primaryClass = "gr-qc",
    doi = "10.1103/PhysRevD.88.064046",
    journal = "Phys. Rev. D",
    volume = "88",
    number = "6",
    pages = "064046",
    year = "2013"
}

@article{Yoshida:1994xi,
    author = "Yoshida, S. and Eriguchi, Y. and Futamase, T.",
    title = "{Quasinormal modes of boson stars}",
    doi = "10.1103/PhysRevD.50.6235",
    journal = "Phys. Rev. D",
    volume = "50",
    pages = "6235--6246",
    year = "1994"
}

@article{Krishnendu:2017shb,
    author = "Krishnendu, N. V. and Arun, K. G. and Mishra, Chandra Kant",
    title = "{Testing the binary black hole nature of a compact binary coalescence}",
    eprint = "1701.06318",
    archivePrefix = "arXiv",
    primaryClass = "gr-qc",
    doi = "10.1103/PhysRevLett.119.091101",
    journal = "Phys. Rev. Lett.",
    volume = "119",
    number = "9",
    pages = "091101",
    year = "2017"
}

@article{Krishnendu:2018nqa,
    author = "Krishnendu, N. V. and Mishra, Chandra Kant and Arun, K. G.",
    title = "{Spin-induced deformations and tests of binary black hole nature using third-generation detectors}",
    eprint = "1811.00317",
    archivePrefix = "arXiv",
    primaryClass = "gr-qc",
    doi = "10.1103/PhysRevD.99.064008",
    journal = "Phys. Rev. D",
    volume = "99",
    number = "6",
    pages = "064008",
    year = "2019"
}

@article{Johnson-Mcdaniel:2018cdu,
    author = "Johnson-Mcdaniel, Nathan K. and Mukherjee, Arunava and Kashyap, Rahul and Ajith, Parameswaran and Del Pozzo, Walter and Vitale, Salvatore",
    title = "{Constraining black hole mimickers with gravitational wave observations}",
    eprint = "1804.08026",
    archivePrefix = "arXiv",
    primaryClass = "gr-qc",
    reportNumber = "LIGO-P1800092",
    doi = "10.1103/PhysRevD.102.123010",
    journal = "Phys. Rev. D",
    volume = "102",
    pages = "123010",
    year = "2020"
}

@article{Chia:2023tle,
    author = "Chia, Horng Sheng and Edwards, Thomas D. P. and Wadekar, Digvijay and Zimmerman, Aaron and Olsen, Seth and Roulet, Javier and Venumadhav, Tejaswi and Zackay, Barak and Zaldarriaga, Matias",
    title = "{In pursuit of Love numbers: First templated search for compact objects with large tidal deformabilities in the LIGO-Virgo data}",
    eprint = "2306.00050",
    archivePrefix = "arXiv",
    primaryClass = "gr-qc",
    doi = "10.1103/PhysRevD.110.063007",
    journal = "Phys. Rev. D",
    volume = "110",
    number = "6",
    pages = "063007",
    year = "2024"
}

@article{Ghosh:2017gfp,
    author = "Ghosh, Abhirup and Johnson-Mcdaniel, Nathan K. and Ghosh, Archisman and Mishra, Chandra Kant and Ajith, Parameswaran and Del Pozzo, Walter and Berry, Christopher P. L. and Nielsen, Alex B. and London, Lionel",
    title = "{Testing general relativity using gravitational wave signals from the inspiral, merger and ringdown of binary black holes}",
    eprint = "1704.06784",
    archivePrefix = "arXiv",
    primaryClass = "gr-qc",
    reportNumber = "LIGO-P1700006, ICTS-2017-3",
    doi = "10.1088/1361-6382/aa972e",
    journal = "Class. Quant. Grav.",
    volume = "35",
    number = "1",
    pages = "014002",
    year = "2018"
}

@article{Ghosh:2016qgn,
    author = "Ghosh, Abhirup and others",
    title = "{Testing general relativity using golden black-hole binaries}",
    eprint = "1602.02453",
    archivePrefix = "arXiv",
    primaryClass = "gr-qc",
    reportNumber = "LIGO-P1500185-V10, ICTS-2016-1, LIGO-P1500185-V11",
    doi = "10.1103/PhysRevD.94.021101",
    journal = "Phys. Rev. D",
    volume = "94",
    number = "2",
    pages = "021101",
    year = "2016"
}

@article{Hofmann:2016yih,
    author = "Hofmann, Fabian and Barausse, Enrico and Rezzolla, Luciano",
    title = "{The final spin from binary black holes in quasi-circular orbits}",
    eprint = "1605.01938",
    archivePrefix = "arXiv",
    primaryClass = "gr-qc",
    doi = "10.3847/2041-8205/825/2/L19",
    journal = "Astrophys. J. Lett.",
    volume = "825",
    number = "2",
    pages = "L19",
    year = "2016"
}

@article{Healy:2016lce,
    author = "Healy, James and Lousto, Carlos O.",
    title = "{Remnant of binary black-hole mergers: New simulations and peak luminosity studies}",
    eprint = "1610.09713",
    archivePrefix = "arXiv",
    primaryClass = "gr-qc",
    doi = "10.1103/PhysRevD.95.024037",
    journal = "Phys. Rev. D",
    volume = "95",
    number = "2",
    pages = "024037",
    year = "2017"
}

@article{Jimenez-Forteza:2016oae,
    author = {Jim\'enez-Forteza, Xisco and Keitel, David and Husa, Sascha and Hannam, Mark and Khan, Sebastian and P\"urrer, Michael},
    title = "{Hierarchical data-driven approach to fitting numerical relativity data for nonprecessing binary black holes with an application to final spin and radiated energy}",
    eprint = "1611.00332",
    archivePrefix = "arXiv",
    primaryClass = "gr-qc",
    reportNumber = "LIGO-P1600270",
    doi = "10.1103/PhysRevD.95.064024",
    journal = "Phys. Rev. D",
    volume = "95",
    number = "6",
    pages = "064024",
    year = "2017"
}

@article{Scheel:2008rj,
  author = "Scheel, M. A. and Boyle, M. and Chu, T. and Kidder, L. E. and Matthews, K. D. and Pfeiffer, H. P.",
  title = "{High-accuracy waveforms for binary black hole inspiral, merger, and ringdown}",
  eprint = "0810.1767",
  archivePrefix = "arXiv",
  primaryClass = "gr-qc",
  doi = "10.1103/PhysRevD.79.024003",
  journal = "Phys. Rev. D",
  volume = "79",
  pages = "024003",
  year = "2009"}

@article{Helfer:2021brt,
    author = "Helfer, Thomas and Sperhake, Ulrich and Croft, Robin and Radia, Miren and Ge, Bo-Xuan and Lim, Eugene A.",
    title = "{Malaise and remedy of binary boson-star initial data}",
    eprint = "2108.11995",
    archivePrefix = "arXiv",
    primaryClass = "gr-qc",
    doi = "10.1088/1361-6382/ac53b7",
    journal = "Class. Quant. Grav.",
    volume = "39",
    number = "7",
    pages = "074001",
    year = "2022"
}

@article{Cardoso:2017cfl,
    author = "Cardoso, Vitor and Franzin, Edgardo and Maselli, Andrea and Pani, Paolo and Raposo, Guilherme",
    title = "{Testing strong-field gravity with tidal Love numbers}",
    eprint = "1701.01116",
    archivePrefix = "arXiv",
    primaryClass = "gr-qc",
    doi = "10.1103/PhysRevD.95.084014",
    journal = "Phys. Rev. D",
    volume = "95",
    number = "8",
    pages = "084014",
    year = "2017",
    note = "[Addendum: Phys.Rev.D 95, 089901 (2017)]"
}

@article{Sennett:2017etc,
    author = "Sennett, Noah and Hinderer, Tanja and Steinhoff, Jan and Buonanno, Alessandra and Ossokine, Serguei",
    title = "{Distinguishing Boson Stars from Black Holes and Neutron Stars from Tidal Interactions in Inspiraling Binary Systems}",
    eprint = "1704.08651",
    archivePrefix = "arXiv",
    primaryClass = "gr-qc",
    doi = "10.1103/PhysRevD.96.024002",
    journal = "Phys. Rev. D",
    volume = "96",
    number = "2",
    pages = "024002",
    year = "2017"
}

@article{Mroue:2010re,
    author = "Mroue, Abdul H. and Pfeiffer, Harald P. and Kidder, Lawrence E. and Teukolsky, Saul A.",
    title = "{Measuring orbital eccentricity and periastron advance in quasi-circular black hole simulations}",
    eprint = "1004.4697",
    archivePrefix = "arXiv",
    primaryClass = "gr-qc",
    doi = "10.1103/PhysRevD.82.124016",
    journal = "Phys. Rev. D",
    volume = "82",
    pages = "124016",
    year = "2010"
}

@article{Sanchis-Gual:2022mkk,
    author = "Sanchis-Gual, Nicolas and Calder{\'o}n Bustillo, Juan and Herdeiro, Carlos and Radu, Eugen and Font, Jos{\'e} A. and Leong, Samson H. W. and Torres-Forn{\'e}, Alejandro",
    title = "{Impact of the wavelike nature of Proca stars on their gravitational-wave emission}",
    eprint = "2208.11717",
    archivePrefix = "arXiv",
    primaryClass = "gr-qc",
    doi = "10.1103/PhysRevD.106.124011",
    journal = "Phys. Rev. D",
    volume = "106",
    number = "12",
    pages = "124011",
    year = "2022"
}

@article{Cheung:2023vki,
    author = "Cheung, Mark Ho-Yeuk and Berti, Emanuele and Baibhav, Vishal and Cotesta, Roberto",
    title = "{Extracting linear and nonlinear quasinormal modes from black hole merger simulations}",
    eprint = "2310.04489",
    archivePrefix = "arXiv",
    primaryClass = "gr-qc",
    doi = "10.1103/PhysRevD.109.044069",
    journal = "Phys. Rev. D",
    volume = "109",
    number = "4",
    pages = "044069",
    year = "2024",
    note = "[Erratum: Phys.Rev.D 110, 049902 (2024), Erratum: Phys.Rev.D 112, 049901 (2025)]"
}

@article{Pompili:2025cdc,
    author = "Pompili, Lorenzo and Maggio, Elisa and Silva, Hector O. and Buonanno, Alessandra",
    title = "{Parametrized spin-precessing inspiral-merger-ringdown waveform model for tests of general relativity}",
    eprint = "2504.10130",
    archivePrefix = "arXiv",
    primaryClass = "gr-qc",
    doi = "10.1103/ng8w-98sz",
    journal = "Phys. Rev. D",
    volume = "111",
    number = "12",
    pages = "124040",
    year = "2025"
}

@article{Andrade2021,
  doi = {10.21105/joss.03703},
  url = {https://doi.org/10.21105/joss.03703},
  year = {2021},
  publisher = {The Open Journal},
  volume = {6},
  number = {68},
  pages = {3703},
  author = {Tomas Andrade and Llibert Areste Salo and Josu C. Aurrekoetxea and Jamie Bamber and Katy Clough and Robin Croft and Eloy de Jong and Amelia Drew and Alejandro Duran and Pedro G. Ferreira and Pau Figueras and Hal Finkel and Tiago Fran\c{c}a and Bo-Xuan Ge and Chenxia Gu and Thomas Helfer and Juha Jäykkä and Cristian Joana and Markus Kunesch and Kacper Kornet and Eugene A. Lim and Francesco Muia and Zainab Nazari and Miren Radia and Justin Ripley and Paul Shellard and Ulrich Sperhake and Dina Traykova and Saran Tunyasuvunakool and Zipeng Wang and James Y. Widdicombe and Kaze Wong},
  title = {GRChombo: An adaptable numerical relativity code for fundamental physics},
  journal = {Journal of Open Source Software}
}

@article{Radia:2021smk,
    author = "Radia, Miren and Sperhake, Ulrich and Drew, Amelia and Clough, Katy and Figueras, Pau and Lim, Eugene A. and Ripley, Justin L. and Aurrekoetxea, Josu C. and Fran{\c{c}}a, Tiago and Helfer, Thomas",
    title = "{Lessons for adaptive mesh refinement in numerical relativity}",
    eprint = "2112.10567",
    archivePrefix = "arXiv",
    primaryClass = "gr-qc",
    reportNumber = "KCL-PH-TH/2021-89",
    doi = "10.1088/1361-6382/ac6fa9",
    journal = "Class. Quant. Grav.",
    volume = "39",
    number = "13",
    pages = "135006",
    year = "2022"
}

@article{Clough:2015sqa,
    author = "Clough, Katy and Figueras, Pau and Finkel, Hal and Kunesch, Markus and Lim, Eugene A. and Tunyasuvunakool, Saran",
    title = "{GRChombo : Numerical Relativity with Adaptive Mesh Refinement}",
    eprint = "1503.03436",
    archivePrefix = "arXiv",
    primaryClass = "gr-qc",
    reportNumber = "KCL-PH-TH-2015-40",
    doi = "10.1088/0264-9381/32/24/245011",
    journal = "Class. Quant. Grav.",
    volume = "32",
    number = "24",
    pages = "245011",
    year = "2015"
}

@misc{exozvezda,
title = {{ExoZvezda code;\\ https://github.com/GRTLCollaboration/ExoZvezda}},
year = "2024",
howpublished = "GitHub repository"}

@article{Thompson:2025hhc,
    author = "Thompson, Jonathan E. and Hoy, Charlie and Fauchon-Jones, Edward and Hannam, Mark",
    title = "{Use and interpretation of signal-model indistinguishability measures for gravitational-wave astronomy}",
    eprint = "2506.10530",
    archivePrefix = "arXiv",
    primaryClass = "gr-qc",
    reportNumber = "LIGO-P2500361",
    doi = "10.1103/ddz7-x9zz",
    journal = "Phys. Rev. D",
    volume = "112",
    number = "6",
    pages = "064011",
    year = "2025"
}

@article{McWilliams:2010eq,
    author = "McWilliams, Sean T. and Kelly, Bernard J. and Baker, John G.",
    title = "{Observing mergers of non-spinning black-hole binaries}",
    eprint = "1004.0961",
    archivePrefix = "arXiv",
    primaryClass = "gr-qc",
    doi = "10.1103/PhysRevD.82.024014",
    journal = "Phys. Rev. D",
    volume = "82",
    pages = "024014",
    year = "2010"
}

@article{Hinderer:2007mb,
    author = "Hinderer, Tanja",
    title = "{Tidal Love numbers of neutron stars}",
    eprint = "0711.2420",
    archivePrefix = "arXiv",
    primaryClass = "astro-ph",
    doi = "10.1086/533487",
    journal = "Astrophys. J.",
    volume = "677",
    pages = "1216--1220",
    year = "2008",
    note = "[Erratum: Astrophys.J. 697, 964 (2009)]"
}

@article{Berti:2009kk,
    author = "Berti, Emanuele and Cardoso, Vitor and Starinets, Andrei O.",
    title = "{Quasinormal modes of black holes and black branes}",
    eprint = "0905.2975",
    archivePrefix = "arXiv",
    primaryClass = "gr-qc",
    doi = "10.1088/0264-9381/26/16/163001",
    journal = "Class. Quant. Grav.",
    volume = "26",
    pages = "163001",
    year = "2009"
}

@article{Regge:1957td,
    author = "Regge, Tullio and Wheeler, John A.",
    title = "{Stability of a Schwarzschild singularity}",
    doi = "10.1103/PhysRev.108.1063",
    journal = "Phys. Rev.",
    volume = "108",
    pages = "1063--1069",
    year = "1957"
}

@article{Zerilli:1970se,
    author = "Zerilli, Frank J.",
    title = "{Effective potential for even parity Regge-Wheeler gravitational perturbation equations}",
    doi = "10.1103/PhysRevLett.24.737",
    journal = "Phys. Rev. Lett.",
    volume = "24",
    pages = "737--738",
    year = "1970"
}

@article{Zerilli:1970wzz,
    author = "Zerilli, F. J.",
    title = "{Gravitational field of a particle falling in a schwarzschild geometry analyzed in tensor harmonics}",
    doi = "10.1103/PhysRevD.2.2141",
    journal = "Phys. Rev. D",
    volume = "2",
    pages = "2141--2160",
    year = "1970"
}

@article{LIGOScientific:2021sio,
    author = "Abbott, R. and others",
    collaboration = "LIGO Scientific, VIRGO, KAGRA",
    title = "{Tests of General Relativity with GWTC-3}",
    eprint = "2112.06861",
    archivePrefix = "arXiv",
    primaryClass = "gr-qc",
    reportNumber = "LIGO-P2100275",
    doi = "10.1103/PhysRevD.112.084080",
    journal = "Phys. Rev. D",
    volume = "112",
    number = "8",
    pages = "084080",
    year = "2025"
}

@article{Madekar:2024zdj,
    author = "Madekar, Sakshi Satish and Johnson-McDaniel, Nathan K. and Gupta, Anuradha and Ghosh, Abhirup",
    title = "{A meta inspiral{\textendash}merger{\textendash}ringdown consistency test of general relativity with gravitational wave signals from compact binaries}",
    eprint = "2405.05884",
    archivePrefix = "arXiv",
    primaryClass = "gr-qc",
    doi = "10.1088/1361-6382/adf02a",
    journal = "Class. Quant. Grav.",
    volume = "42",
    number = "16",
    pages = "165011",
    year = "2025"
}

@article{Shaikh:2024wyn,
    author = "Shaikh, Md Arif and Bhat, Sajad A. and Kapadia, Shasvath J.",
    title = "{A study of the inspiral-merger-ringdown consistency test with gravitational-wave signals from compact binaries in eccentric orbits}",
    eprint = "2402.15110",
    archivePrefix = "arXiv",
    primaryClass = "gr-qc",
    doi = "10.1103/PhysRevD.110.024030",
    journal = "Phys. Rev. D",
    volume = "110",
    number = "2",
    pages = "024030",
    year = "2024"
}

@article{Radia:2021hjs,
    author = "Radia, Miren and Sperhake, Ulrich and Berti, Emanuele and Croft, Robin",
    title = "{Anomalies in the gravitational recoil of eccentric black-hole mergers with unequal mass ratios}",
    eprint = "2101.11015",
    archivePrefix = "arXiv",
    primaryClass = "gr-qc",
    doi = "10.1103/PhysRevD.103.104006",
    journal = "Phys. Rev. D",
    volume = "103",
    number = "10",
    pages = "104006",
    year = "2021"
}

@article{Ashton:2018jfp,
    author = "Ashton, Gregory and others",
    title = "{BILBY: A user-friendly Bayesian inference library for gravitational-wave astronomy}",
    eprint = "1811.02042",
    archivePrefix = "arXiv",
    primaryClass = "astro-ph.IM",
    doi = "10.3847/1538-4365/ab06fc",
    journal = "Astrophys. J. Suppl.",
    volume = "241",
    number = "2",
    pages = "27",
    year = "2019"
}

@software{nessai,
  author       = {Michael J. Williams},
  title        = {nessai: Nested Sampling with Artificial Intelligence},
  month        = feb,
  year         = 2021,
  publisher    = {Zenodo},
  version      = {latest},
  doi          = {10.5281/zenodo.4550693},
  url          = {https://doi.org/10.5281/zenodo.4550693}
}

@article{Williams:2021qyt,
    author = "Williams, Michael J. and Veitch, John and Messenger, Chris",
    title = "{Nested sampling with normalizing flows for gravitational-wave inference}",
    eprint = "2102.11056",
    archivePrefix = "arXiv",
    primaryClass = "gr-qc",
    doi = "10.1103/PhysRevD.103.103006",
    journal = "Phys. Rev. D",
    volume = "103",
    number = "10",
    pages = "103006",
    year = "2021"
}

@article{Pratten:2020fqn,
    author = "Pratten, Geraint and Husa, Sascha and Garcia-Quiros, Cecilio and Colleoni, Marta and Ramos-Buades, Antoni and Estelles, Hector and Jaume, Rafel",
    title = "{Setting the cornerstone for a family of models for gravitational waves from compact binaries: The dominant harmonic for nonprecessing quasicircular black holes}",
    eprint = "2001.11412",
    archivePrefix = "arXiv",
    primaryClass = "gr-qc",
    reportNumber = "LIGO-P2000018",
    doi = "10.1103/PhysRevD.102.064001",
    journal = "Phys. Rev. D",
    volume = "102",
    number = "6",
    pages = "064001",
    year = "2020"
}

@article{Pratten:2020ceb,
    author = "Pratten, Geraint and others",
    title = "{Computationally efficient models for the dominant and subdominant harmonic modes of precessing binary black holes}",
    eprint = "2004.06503",
    archivePrefix = "arXiv",
    primaryClass = "gr-qc",
    doi = "10.1103/PhysRevD.103.104056",
    journal = "Phys. Rev. D",
    volume = "103",
    number = "10",
    pages = "104056",
    year = "2021"
}

@article{KAGRA:2013rdx,
    author = "Abbott, B. P. and others",
    collaboration = "KAGRA, LIGO Scientific, Virgo",
    title = "{Prospects for observing and localizing gravitational-wave transients with Advanced LIGO, Advanced Virgo and KAGRA}",
    eprint = "1304.0670",
    archivePrefix = "arXiv",
    primaryClass = "gr-qc",
    reportNumber = "LIGO-P1200087, VIR-0288A-12, JGW-P1808427, LIGO-T2000012-v1",
    doi = "10.1007/s41114-020-00026-9",
    journal = "Living Rev. Rel.",
    volume = "19",
    pages = "1",
    year = "2016"
}

@misc{chombo,
  author = {Adams, M. and others},
  title = "{Chombo Software Package for AMR Applications - Design Document}",
  number= "{LBNL}-6616{E}",
  year = {2019},
  institution = {Lawrence Berkeley National Laboratory}}

@article{Kaup:1968zz,
    author = "Kaup, David J.",
    title = "{Klein-Gordon Geon}",
    doi = "10.1103/PhysRev.172.1331",
    journal = "Phys. Rev.",
    volume = "172",
    pages = "1331--1342",
    year = "1968"
}

@article{Liebling:2012fv,
    author = "Liebling, Steven L. and Palenzuela, Carlos",
    title = "{Dynamical boson stars}",
    eprint = "1202.5809",
    archivePrefix = "arXiv",
    primaryClass = "gr-qc",
    doi = "10.1007/s41114-023-00043-4",
    journal = "Living Rev. Rel.",
    volume = "26",
    number = "1",
    pages = "1",
    year = "2023"
}

@inbook{Bezares:2024btu,
    author = "Bezares, Miguel and Sanchis-Gual, Nicolas",
    title = "{Exotic Compact Objects: A~Recent Numerical-Relativity Perspective}",
    eprint = "2406.04901",
    archivePrefix = "arXiv",
    primaryClass = "gr-qc",
    doi = "10.1007/978-981-97-8522-3_18",
    year = "2025"
}

@article{PhysRevD.35.3658,
  title = {Scalar soliton stars and black holes},
  author = {Friedberg, R. and Lee, T. D. and Pang, Y.},
  journal = {Phys. Rev. D},
  volume = {35},
  issue = {12},
  pages = {3658--3677},
  numpages = {0},
  year = {1987},
  month = {Jun},
  publisher = {American Physical Society},
  doi = {10.1103/PhysRevD.35.3658},
  url = {https://link.aps.org/doi/10.1103/PhysRevD.35.3658}
}

@article{Kleihaus:2005me,
    author = "Kleihaus, Burkhard and Kunz, Jutta and List, Meike",
    title = "{Rotating boson stars and Q-balls}",
    eprint = "gr-qc/0505143",
    archivePrefix = "arXiv",
    doi = "10.1103/PhysRevD.72.064002",
    journal = "Phys. Rev. D",
    volume = "72",
    pages = "064002",
    year = "2005"
}

@article{Palenzuela:2006wp,
    author = "Palenzuela, C. and Olabarrieta, I. and Lehner, L. and Liebling, Steven L.",
    title = "{Head-on collisions of boson stars}",
    eprint = "gr-qc/0612067",
    archivePrefix = "arXiv",
    doi = "10.1103/PhysRevD.75.064005",
    journal = "Phys. Rev. D",
    volume = "75",
    pages = "064005",
    year = "2007"
}

@article{LIGOScientific:2025slb,
    author = "Abac, A. G. and others",
    collaboration = "LIGO Scientific, VIRGO, KAGRA",
    title = "{GWTC-4.0: Updating the Gravitational-Wave Transient Catalog with Observations from the First Part of the Fourth LIGO-Virgo-KAGRA Observing Run}",
    eprint = "2508.18082",
    journal = "",
    archivePrefix = "arXiv",
    primaryClass = "gr-qc",
    reportNumber = "LIGO-P2400386",
    month = "8",
    year = "2025"
}

@article{Bezares:2022obu,
    author = "Bezares, Miguel and Bo{\v{s}}kovi{\'c}, Mateja and Liebling, Steven and Palenzuela, Carlos and Pani, Paolo and Barausse, Enrico",
    title = "{Gravitational waves and kicks from the merger of unequal mass, highly compact boson stars}",
    eprint = "2201.06113",
    archivePrefix = "arXiv",
    primaryClass = "gr-qc",
    doi = "10.1103/PhysRevD.105.064067",
    journal = "Phys. Rev. D",
    volume = "105",
    number = "6",
    pages = "064067",
    year = "2022"
}

@article{Ge:2024itl,
    author = "Ge, Bo-Xuan and Lim, Eugene A. and Sperhake, Ulrich and Evstafyeva, Tamara and Cors, Daniela and de Jong, Eloy and Croft, Robin and Helfer, Thomas",
    title = "{Dynamics and gravitational radiation of stable and unstable boson-star mergers}",
    eprint = "2410.23839",
    archivePrefix = "arXiv",
    primaryClass = "gr-qc",
    reportNumber = "KCL-TH-2024-59",
    doi = "10.1103/2dhs-phl4",
    journal = "Phys. Rev. D",
    volume = "112",
    number = "12",
    pages = "124080",
    year = "2025"
}

@article{Evstafyeva:2022bpr,
    author = "Evstafyeva, Tamara and Sperhake, Ulrich and Helfer, Thomas and Croft, Robin and Radia, Miren and Ge, Bo-Xuan and Lim, Eugene A.",
    title = "{Unequal-mass boson-star binaries: initial data and merger dynamics}",
    eprint = "2212.08023",
    archivePrefix = "arXiv",
    primaryClass = "gr-qc",
    doi = "10.1088/1361-6382/acc2a8",
    journal = "Class. Quant. Grav.",
    volume = "40",
    number = "8",
    pages = "085009",
    year = "2023"
}

@article{Siemonsen:2023age,
    author = "Siemonsen, Nils and East, William E.",
    title = "{Generic initial data for binary boson stars}",
    eprint = "2306.17265",
    archivePrefix = "arXiv",
    primaryClass = "gr-qc",
    doi = "10.1103/PhysRevD.108.124015",
    journal = "Phys. Rev. D",
    volume = "108",
    number = "12",
    pages = "124015",
    year = "2023"
}

@article{Bezares:2017mzk,
    author = "Bezares, Miguel and Palenzuela, Carlos and Bona, Carles",
    title = "{Final fate of compact boson star mergers}",
    eprint = "1705.01071",
    archivePrefix = "arXiv",
    primaryClass = "gr-qc",
    doi = "10.1103/PhysRevD.95.124005",
    journal = "Phys. Rev. D",
    volume = "95",
    number = "12",
    pages = "124005",
    year = "2017"
}

@article{Leaver:1985ax,
    author = "Leaver, E. W.",
    title = "{An Analytic representation for the quasi normal modes of Kerr black holes}",
    doi = "10.1098/rspa.1985.0119",
    journal = "Proc. Roy. Soc. Lond. A",
    volume = "402",
    pages = "285--298",
    year = "1985"
}

@article{Shibata:1995we,
  author = "Shibata, M. and Nakamura, T.",
  title = "Evolution of three-dimensional gravitational waves: {H}armonic slicing case",
  journal = "Phys. Rev. D",
  year = "1995",
  doi = "10.1103/PhysRevD.52.5428",
  volume = "52",
  pages = "5428-5444"}

@article{Baumgarte:1998te,
  author = "Baumgarte, T. W. and Shapiro, S. L.",
  title = "On the {N}umerical integration of {E}instein's field equations",
  journal = "Phys. Rev. D",
  year = "1998",
  doi = "10.1103/PhysRevD.59.024007",
  volume = "59",
  pages = "024007",
  note = "gr-qc/9810065"}

@article{Sperhake:2006cy,
  author = "Sperhake, U.",
  title = "Binary black-hole evolutions of excision and puncture data",
  year = "2007",
  journal = "Phys. Rev. D",
  volume = "76",
  pages = "104015",
  doi = "10.1103/PhysRevD.76.104015",
  note = "gr-qc/0606079"}

@misc{Allen:1999,
  author = "Allen, G. and Goodale, T. and Mass\'o, J. and Seidel, E.",
  title = "{The Cactus Computational Toolkit and Using Distributed Computing to Collide Neutron Stars}",
  booktitle = "{Proceedings of Eighth IEEE International Symposium on High Performance Distributed Computing, HPDC-8, Redondo Beach, 1999}",
  address = "{Piscataway, New Jersey, United States}",
  publisher = "{IEEE Press}",
  year = "1999"}

@article{Schnetter:2003rb,
  author = "Schnetter, E. and Hawley, S. H. and Hawke, I.",
  title = "{Evolutions in 3-D numerical relativity using fixed mesh refinement}",
  journal = "Class. Quant. Grav.",
  volume = "21",
  pages = "1465-1488",
  doi = "10.1088/0264-9381/21/6/014",
  year = "2004",
  note = "gr-qc/0310042"}

@article{Thornburg:1995cp,
  author = "Thornburg, J.",
  title = "{Finding apparent horizons in numerical relativity}",
  journal = "Phys. Rev. D",
  volume = "54",
  pages = "4899-4918",
  doi = "10.1103/PhysRevD.54.4899",
  year = "1996",
  note = "gr-qc/9508014"}

@article{Thornburg:2003sf,
  author = "Thornburg, J.",
  title = "{A Fast apparent horizon finder for three-dimensional Cartesian grids in numerical relativity}",
  journal = "Class. Quant. Grav.",
  volume = "21",
  pages = "743-766",
  doi = "10.1088/0264-9381/21/2/026",
  year = "2004",
  note = "gr-qc/0306056"}

@article{Buonanno:1998gg,
    author = "Buonanno, A. and Damour, T.",
    title = "{Effective one-body approach to general relativistic two-body dynamics}",
    eprint = "gr-qc/9811091",
    archivePrefix = "arXiv",
    reportNumber = "IHES-P-98-74",
    doi = "10.1103/PhysRevD.59.084006",
    journal = "Phys. Rev. D",
    volume = "59",
    pages = "084006",
    year = "1999"
}

@article{Mills:2020thr,
    author = "Mills, Cameron and Fairhurst, Stephen",
    title = "{Measuring gravitational-wave higher-order multipoles}",
    eprint = "2007.04313",
    archivePrefix = "arXiv",
    primaryClass = "gr-qc",
    doi = "10.1103/PhysRevD.103.024042",
    journal = "Phys. Rev. D",
    volume = "103",
    number = "2",
    pages = "024042",
    year = "2021"
}

@article{Damour:2025oys,
    author = "Damour, Thibault and Jain, Tamanna and Sperhake, Ulrich",
    title = "{Gravitational scattering of solitonic boson stars: Analytics vs Numerics}",
    eprint = "2512.00945",
    archivePrefix = "arXiv",
    primaryClass = "gr-qc",
    journal = "",
    month = "11",
    year = "2025"
}

@article{Duez:2004nf,
  author = "Duez, M. D. and Liu, Y. T. and Shapiro, S. L. and Stephens, B. C.",
  title = "{General relativistic hydrodynamics with viscosity: Contraction, catastrophic collapse, and disk formation in hypermassive neutron stars}",
  eprint = "astro-ph/0402502",
  archivePrefix = "arXiv",
  doi = "10.1103/PhysRevD.69.104030",
  journal = "Phys. Rev. D",
  volume = "69",
  pages = "104030",
  year = "2004"}

@article{LIGOScientific:2025rid,
  author = "Abac, A. G. and others",
  collaboration = "LIGO Scientific, Virgo, KAGRA",
  title = "{GW250114: Testing Hawking{\textquoteright}s Area Law and the Kerr Nature of Black Holes}",
  eprint = "2509.08054",
  archivePrefix = "arXiv",
  primaryClass = "gr-qc",
  reportNumber = "LIGO-P2500421",
  doi = "10.1103/kw5g-d732",
  journal = "Phys. Rev. Lett.",
  volume = "135",
  number = "11",
  pages = "111403",
  year = "2025"}

@article{Scheel:2025jct,
    author = "Scheel, Mark A. and others",
    title = "{The SXS collaboration{\textquoteright}s third catalog of binary black hole simulations}",
    eprint = "2505.13378",
    archivePrefix = "arXiv",
    primaryClass = "gr-qc",
    doi = "10.1088/1361-6382/adfd34",
    journal = "Class. Quant. Grav.",
    volume = "42",
    number = "19",
    pages = "195017",
    year = "2025"
}

@article{Toubiana:2024car,
    author = "Toubiana, Alexandre and Gair, Jonathan R.",
    title = "{Indistinguishability criterion and estimating the presence of biases}",
    journal = "arXiv",
    eprint = "2401.06845",
    archivePrefix = "arXiv",
    primaryClass = "gr-qc",
    month = "1",
    year = "2024"
}

@article{Sperhake:2011zz,
  author = "Sperhake, U. and Br{\"u}gmann, B. and M{\"u}ller, D. and Sopuerta, C. F.",
  title = "{11-orbit inspiral of a mass ratio 4:1 black-hole binary}",
  journal = "Class. Quant. Grav.",
  volume = "28",
  pages = "134004",
  year = "2011",
  doi = "10.1088/0264-9381/28/13/134004",
  note = "arXiv:1012.3173 [gr-qc]"}

@article{Ajith:2007kx,
    author = "Ajith, P. and others",
    title = "{A Template bank for gravitational waveforms from coalescing binary black holes. I. Non-spinning binaries}",
    eprint = "0710.2335",
    archivePrefix = "arXiv",
    primaryClass = "gr-qc",
    reportNumber = "LIGO-P070111-00-Z, AEI-2007-143",
    doi = "10.1103/PhysRevD.77.104017",
    journal = "Phys. Rev. D",
    volume = "77",
    pages = "104017",
    year = "2008",
    note = "[Erratum: Phys.Rev.D 79, 129901 (2009)]"
}

@article{Ajith:2007qp,
    author = "Ajith, Parameswaran and others",
    editor = "Krishnan, B. and Papa, M. A. and Schutz, Bernard F.",
    title = "{Phenomenological template family for black-hole coalescence waveforms}",
    eprint = "0704.3764",
    archivePrefix = "arXiv",
    primaryClass = "gr-qc",
    doi = "10.1088/0264-9381/24/19/S31",
    journal = "Class. Quant. Grav.",
    volume = "24",
    pages = "S689--S700",
}

@article{LIGOScientific:2026qni,
    author = "Abac, A. G. and others",
    collaboration = "LIGO Scientific, VIRGO, KAGRA",
    title = "{GWTC-4.0: Tests of General Relativity. I. Overview and General Tests}",
    eprint = "2603.19019",
    archivePrefix = "arXiv",
    journal="",
    primaryClass = "gr-qc",
    reportNumber = "LIGO-P2500065",
    month = "3",
    year = "2026"
}

@article{LIGOScientific:2026fcf,
    author = "Abac, A. G. and others",
    collaboration = "LIGO Scientific, VIRGO, KAGRA",
    title = "{GWTC-4.0: Tests of General Relativity. II. Parameterized Tests}",
    eprint = "2603.19020",
    archivePrefix = "arXiv",
    journal="",
    primaryClass = "gr-qc",
    reportNumber = "LIGO-P2500066",
    month = "3",
    year = "2026"
}

@article{LIGOScientific:2026wpt,
    author = "Abac, A. G. and others",
    collaboration = "LIGO Scientific, VIRGO, KAGRA",
    title = "{GWTC-4.0: Tests of General Relativity. III. Tests of the Remnants}",
    eprint = "2603.19021",
    archivePrefix = "arXiv",
    journal="",
    primaryClass = "gr-qc",
    reportNumber = "LIGO-P2500067",
    month = "3",
    year = "2026"
}

@article{Puecher:2022sfm,
    author = "Puecher, Anna and Kalaghatgi, Chinmay and Roy, Soumen and Setyawati, Yoshinta and Gupta, Ish and Sathyaprakash, B. S. and Van Den Broeck, Chris",
    title = "{Testing general relativity using higher-order modes of gravitational waves from binary black holes}",
    eprint = "2205.09062",
    archivePrefix = "arXiv",
    primaryClass = "gr-qc",
    reportNumber = "LIGO DCC: LIGO-P2200140",
    doi = "10.1103/PhysRevD.106.082003",
    journal = "Phys. Rev. D",
    volume = "106",
    number = "8",
    pages = "082003",
    year = "2022"
}

@article{Gupta:2025paz,
    author = "Gupta, Ish and Narayan, Purnima and London, Lionel and Tiwari, Shubhanshu and Sathyaprakash, Bangalore",
    title = "{Testing general relativity with amplitudes of subdominant gravitational-wave modes}",
    journal = "",
    eprint = "2511.11886",
    archivePrefix = "arXiv",
    primaryClass = "gr-qc",
    month = "11",
    year = "2025"
}

@article{Chiaramello:2025bhi,
    author = "Chiaramello, Danilo and Cibrario, Nicol{\`o} and Lange, Jacob and Chandra, Koustav and Gamba, Rossella and Bonino, Raffaella and Nagar, Alessandro",
    title = "{A parametrized model for gravitational waves from eccentric, precessing binary black holes: theory-agnostic tests of General Relativity with pTEOBResumS}",
    eprint = "2511.19593",
    journal = "",
    archivePrefix = "arXiv",
    primaryClass = "gr-qc",
    month = "11",
    year = "2025"
}

@article{Johnson-McDaniel:2021yge,
    author = "Johnson-McDaniel, Nathan K. and Ghosh, Abhirup and Ghonge, Sudarshan and Saleem, Muhammed and Krishnendu, N. V. and Clark, James A.",
    title = "{Investigating the relation between gravitational wave tests of general relativity}",
    eprint = "2109.06988",
    archivePrefix = "arXiv",
    primaryClass = "gr-qc",
    reportNumber = "LIGO-P2100322",
    doi = "10.1103/PhysRevD.105.044020",
    journal = "Phys. Rev. D",
    volume = "105",
    number = "4",
    pages = "044020",
    year = "2022"
}

\end{document}